\documentclass{elsarticle}

\usepackage{amssymb}
\usepackage{graphicx}
\usepackage{epstopdf, epsfig}
\usepackage{lineno,hyperref}
\usepackage{booktabs} 
\usepackage{array} 
\usepackage{amsmath}
\usepackage{mathtools}
\usepackage{graphicx,caption,subcaption,color}

\journal{Journal of \LaTeX\ Templates}

\begin{document}

\begin{frontmatter}

\title{A thermodynamically consistent pseudo-potential lattice Boltzmann model for multi-component, multiphase, partially miscible mixtures}

\author[label1]{Cheng Peng\corref{cor1}}
\ead{czp341@psu.edu}

\author[label1]{Luis F. Ayala}
\author[label2]{Orlando M. Ayala}

\address[label1]{Department of Energy and Mineral Engineering and EMS Energy Institute,
The Pennsylvania State University,
University Park, PA 16802, USA}
\address[label2]{111A Kaufman Hall, Department of Engineering Technology, Old Dominion University, Norfolk, VA, 23529, USA}

\cortext[cor1]{Corresponding author}

\begin{abstract}
Current multi-component, multiphase pseudo-potential lattice Boltzmann models have thermodynamic inconsistencies that prevent them to correctly predict the thermodynamic phase behavior of partially miscible multi-component mixtures, such as hydrocarbon mixtures.  This paper identifies these inconsistencies and attempts to design a thermodynamically consistent multi-component, multiphase pseudo-potential lattice Boltzmann model that allows mass transfer across the phase interfaces and is capable to predict the phase behavior of typically partially miscible hydrocarbon mixtures. The designed model defines the total interaction force for the entire phase and split the force into individual components. Through a properly derived force split factor associated with the volatility of each component, the model can achieve precise thermodynamic consistency in multi-component hydrocarbon mixtures, which is described by the iso-fugacity rule.
\end{abstract}


\end{frontmatter}

\section{Introduction}

In the oil and gas industry, there has always been a need for reliable numerical tools able to conduct direct numerical investigations for pore-scale multi-component multi-phase flows in porous media to understand and predict the fluid behaviors in conventional and unconventional reservoirs~\cite{ertekin2018reservoir}. 
Oil and natural gases are multi-component hydrocarbon mixtures that behave as highly non-ideal fluids that undergo phase transitions under pressure, temperature and composition changes~\cite{ertekin2018reservoir,danesh1998pvt}. When such a multi-component mixture separates into two phases, {\it i.e.}, liquid and vapor phases, not only each phase may contain all components, but also the composition of those components in each phase will be often very different. Within each phase, molecules form a homogeneous (fully miscible) mixture with thermodynamic properties that are significantly different from those of pure phases containing only one component~\cite{danesh1998pvt}. Accounting for, and allowing mass transfer of every component across interfaces is a necessity in these types of thermodynamic systems.

Over the last two decades, multiphase (MP) lattice Boltzmann (LB) models have been developed and deployed in a variety of applications. The pseudo-potential (PP) LB models, also known as Shan-Chen models, are among one of the most popular categories of multiphase LB models, due to their conceptual simplicity and numerical efficiency~\cite{shan1993lattice,shan1994simulation}. The PP LB models define external body forces, known as Shan-Chen forces that mimic the molecular interactions to achieve phase separation. There are many variations of PP LB models for multi-component (MC) applications~\cite{martys1996simulation,shan1995multicomponent,shan1996diffusion,yu2011numerical,bao2013lattice,chen2015pore}. 
In those models, each individual component was assigned with a set of distribution functions, and a Shan-Chen force was defined to control its diffusion into other components. Through a Chapman-Enskog analysis, Shan and Doolen showed that the mutual diffusivities of a two-component system were controlled by the Shan-Chen forces and the relaxation times in the lattice Boltzmann equation (LBE), adjusting which the desired diffusivities could be obtained~\cite{shan1996diffusion}. By assigning different mutual diffusivities, authors suggested that these MC models could be applied to components with various miscibilities. However, as we shall see later in Sec.~\ref{sec:LBM}, this way of directly defining Shan-Chen forces in terms of individual components is not thermodynamic consistent and triggers thermodynamic problems in multi-component systems in a number of ways described in this document. 

Another typical constraint in the available MCMP PP LB models is the difficulty to achieve large density and viscosity ratios between phases. So far, the most successful application of these available MCMP PP LB models is the study of oil-water-type flow in porous media and microchannels~\cite{pan2004lattice,li2005pore}. For such applications, each component can be treated as an incompressible fluid with close densities and viscosities, and repulsive Shan-Chen forces were assigned to prevent different components from diffusing into each other. There are several recent attempts ({\it e.g.}~\cite{yu2011numerical,bao2013lattice,kamali2013simulating,chen2015pore,stiles2016high}) to combine the single-component (SC) PP LB models with MC PP models to achieve higher density ratios, as reviewed by Chen {\it et al}.~\cite{chen2014critical}. These attempts usually target applications such as air bubbles rising in water environment or water droplets appearing in air environment, where water and air are fully immiscible, which disallowing mass transfer across the interface. These applications do not incorporate the most important features of multi-component thermodynamics.

There have been attempts to study the flow of miscible fluids in the LB literature. Some of them are constrained to single-phase flow scenarios. For example, in the early work of Flekk{\o}y~\cite{flekkoy1993lattice} and the recent work of Meng and Guo~\cite{meng2015multiple}, ``miscible fluids" correspond to solvents. LB models were developed to study how the transport of a solute through a solvent was affected by the flow of the solvent. This type of single-phase LB models is not applicable to multiphase scenarios. Luo and Girimaji~\cite{luo2003theory} and later Zhu {\it et al.}~\cite{zhu2006simulation} developed two-fluid LB models that introduced extra collision terms in additional to the regular collision terms in LBE to handle the mutual diffusivities between two fluids. Compared to MC PP LB models, this type of models is better rooted to the Boltzmann equation and resulting mutual diffusivities solely depend on the relaxation times of the extra collision terms~\cite{luo2003theory}. Through the adjustment of mutual diffusivities in the two-fluid system, the authors state the entire spectrum of full miscibility to full immiscibility may be achieved. However, multi-component thermodynamics considerations are still absent from these models. Therefore, these methods are unlikely to be able to achieve thermodynamic consistency when applied to partially miscible multi-component hydrocarbon mixtures due to the absence of such information.

Our main targets in this document are: 1) discussing the thermodynamic inconsistencies in the previous MCMP PP LB models on predicting the phase behavior of partially miscible hydrocarbon mixtures, and 2) developing a correct model for this purpose and contrasting it against other recent efforts. In the developed model, we will define the total interaction force in terms of bulk phases and split it into components, and create a force split factor that is consistent with the PP LB models by rigorously deriving its definition. The developed model allows each phase to not only contain all components in the system, but also have fluid properties that are fully consistent with the MC thermodynamics.

The remaining of this document is arranged as follows. In Sec.~\ref{sec:LBM}, we will briefly introduce the MCMP PP LB models for immiscible components and discuss their inadequacies for applications with miscible MC mixtures. Then, a correct model will be introduced in Sec.~\ref{sec:model}. Particularly, the design of a proper force split factor to satisfy precise thermodynamic consistency will be discussed in detail. The proposed model will be validated in four cases, a limiting case with components being identical and three general cases of two-component propane and pentane hydrocarbon mixtures in Sec.~\ref{sec:validation}. Conclusions and final remarks will be summarized in Sec.~\ref{sec:conclusion}. 

\section{Thermodynamic inconsistencies in MCMP PP LB models}\label{sec:LBM}

\subsection{MCMP PP LB models}
In MCMP PP LB models, each component is assigned with a separate set of particle distribution functions, whose evolution is described by LBE as~\cite{he1997theory,guo2002discrete}
\begin{equation}
f_{\sigma,\alpha}\left({\bf x}+{\bf e}_{\alpha}\delta_{t},t+\delta_{t}\right) - f_{\sigma,\alpha}\left({\bf x},t\right)  = -\frac{1}{\tau}\left[f_{\sigma,\alpha}\left({\bf x},t\right)-f_{\sigma,\alpha}^{(eq)}\left({\bf x},t\right)\right] + F_{\sigma,\alpha}\left({\bf x},t\right),
    \label{eq:LBE}
\end{equation}
where $f_{\sigma,\alpha}$ is the particle distribution function of component $\sigma$ that is associated with the particle velocity ${\bf e}_{\alpha}$, ${\bf x}$ and $t$ are the spatial and temporal coordinates, respectively, $\delta t$ is the time step size, $\tau$ is the non-dimensional relaxation time that is related to the kinematic viscosity $\nu$ of the fluid as $\tau = \nu/\left(c_s^2\delta t\right) + 0.5$. $f_{\sigma,\alpha}^{(eq)}$ is the equilibrium distribution of $f_{\sigma,\alpha}$, $F_{\sigma,\alpha}$ is the forcing term representing the effect of external body force on the evolution of $f_{\sigma,\alpha}$. These two terms can be defined in terms of the macroscopic properties as
~\cite{guo2002discrete} 
\begin{subequations}
\begin{align}
    &f_{\sigma,\alpha}^{(eq)} = \bar{\rho}_{\sigma} w_{\alpha} \left[1+\frac{{\bf e}_{\alpha } \cdot {\bf u}}{c_s^2} + \frac{\left( {\bf e}_{\alpha}\cdot {\bf u}\right)^{2}}{2c_s^4} -  \frac{{\bf u}\cdot{\bf u}}{2c_s^2}\right],\\
    &F_{\sigma,\alpha} = \left(1-\frac{1}{2\tau}\right)w_{\alpha}\left[\frac{ {\bf e}_{\alpha} - {\bf u}}{c_s^2} + \frac{ \left( {\bf e}_{\alpha}\cdot {\bf u}\right){\bf e}_{\alpha}}{c_s^4}\right]\cdot {\bf F}_{\sigma}\delta t,
    \end{align}
    \label{eq:equilibrium}
\end{subequations}
where $w_{\alpha}$ is the weighting factor, $\bar{\rho}_{\sigma}$ is the \emph{cell-volume density} of component $\sigma$, {\it i.e.}, the mass of component $\sigma$ per the control volume of a grid cell, 
${\bf u}$ and ${\bf F}_{\sigma}$ are the fluid velocity, and the external body force acting on the component $\sigma$, respectively.
Through the Chapman-Enskog multiscale expansion, the Navier-Stoke equations (NSE) can be recovered from above LBE~\cite{huang2015multiphase}, {\it i.e.}, 
\begin{subequations}
\begin{align}
&\frac{\partial \rho}{\partial t} + \boldsymbol{\nabla}\cdot\left(\rho {\bf u}\right) = 0,\\
&\frac{\partial }{\partial t}\left(\rho {\bf u}\right) + \boldsymbol{\nabla}\cdot\left(\rho {\bf u}{\bf u}\right) = \boldsymbol{\nabla}\cdot\left(-p{\bf I} + {\bf T}\right) + {\bf F}.
\end{align}
    \label{eq:NS}
\end{subequations}
In Eq.~(\ref{eq:NS}), $\rho$ without an overbar is the thermodynamic (bulk) density of the fluid or the density of the phase, which is the summation of all the cell-volume densities for individual components, $p$ is the pressure, ${\bf I}$ is a unit matrix, ${\bf T}$ is the viscous stress tensor, and ${\bf F}$ is the total external body force.

In standard LB models, the pressure $p$ and the fluid density $\rho$ are linearly coupled, which does not reflect the non-ideal pressure-density relationship that triggers phase transitions.
PP LB models achieve multiphase flow simulations via adding external body forces, usually referred as Shan-Chen forces, to NSE to mimic the macroscopic effects of molecular interactions~\cite{shan1993lattice,shan1994simulation}. In MCMP PP LB models, such Shan-Chen forces are applied to individual components~\cite{martys1996simulation,huang2015multiphase}. 
A representative design of Shan-Chen forces for MC systems that allows high density ratios between two phases was given by Bao and Schaefer~\cite{bao2013lattice}. In this work, the total Shan-Chen force acting on the component $\sigma$ is formulated as~\cite{bao2013lattice}
\begin{equation}
\begin{split}
    {\bf F}_{\sigma}\left({\bf x}\right) =& -g_{\sigma\sigma}\psi_{\sigma}\left({\bf x}\right)\sum_{\alpha}w_{\alpha}\psi_{\sigma}\left({\bf x}+{\bf e}_{\alpha}\right){\bf e}_{\alpha}\\
    & -g_{\sigma\bar{\sigma}}\psi_{\sigma}\left({\bf x}\right)\sum_{\alpha}w_{\alpha}\psi_{\bar{\sigma}}\left({\bf x}+{\bf e}_{\alpha}\right){\bf e}_{\alpha},
     \end{split}
    \label{eq:MCMPforce}
\end{equation}
where the first term is the intra-molecular interaction force within component $\sigma$, and the second term is the inter-component interaction force (assuming a binary system) with another component $\bar{\sigma}$. $g_{\sigma\sigma}$ and $g_{\sigma{\bar{\sigma}}}$ are the forcing intensities of the two parts of the force, $\psi_{\sigma}$ and $\psi_{\bar{\sigma}}$ are the pseudo-potential or effective mass of the component $\sigma$ and $\bar{\sigma}$, respectively. ${\bf x}$ and ${\bf x} + {\bf e_{\alpha}}$ are the spatial location of the current location and its neighboring location, ${\bf e_{\alpha}}$ is the $\alpha$th lattice direction. The intra-molecular interaction in Eq.~(\ref{eq:MCMPforce}) serves two purposes. First, it is expected to reproduce the pressure of the non-ideal fluid in each phase. Second, it is found to help achieving a higher density ratio between different phases. It should be emphasized that in these previous works, terms ``component" and ``phase'' were used interchangeably because each component was allowed to be present in only one phase\footnote{This assumes full immiscibility of each component. In reality, each phase is more likely to contain at least trace amounts of all components present in the system.}. Rigorously speaking, the concept of thermodynamic (bulk) density applies to phases only. Components in a phase do not have thermodynamic meaningful density. 

To implement Eq.~(\ref{eq:MCMPforce}), one must define the two effective masses $\psi_{\sigma}$ and $\psi_{\bar{\sigma}}$. Bao and Schaefer~\cite{bao2013lattice} suggested that $\psi_{\sigma}$ and $\psi_{\bar{\sigma}}$ could be calculated using the equations of state (EOS) of \emph{pure substances} as 
\begin{equation}
\psi_{\sigma} = \sqrt{\frac{2\left[p_{EOS}\left(\bar{\rho}_{\sigma}\right) - c_s^2\rho_{\sigma}\right]}{g_{\sigma\sigma}c_s^{2}}},~~~~\psi_{\bar{\sigma}} = \sqrt{\frac{2\left[p'_{EOS'}\left(\bar{\rho}_{\bar{\sigma}}\right) - c_s^2\rho_{\bar{\sigma}}\right]}{g_{\bar{\sigma}\bar{\sigma}}c_s^{2}}},
    \label{eq:effectivemass}
\end{equation}
where $g_{\bar{\sigma}\bar{\sigma}}$ is the forcing intensity of the other component $\bar{\sigma}$, $p_{EOS}$ and $p'_{EOS'}$ are the thermodynamic pressure of component $\sigma$ and $\bar{\sigma}$ calculated from the thermodynamic EOS, {\it e.g.}, Peng-Robinson EOS~\cite{peng1976new}
\begin{equation}
\begin{split}
&p_{EOS}\left(\rho\right) = \frac{\rho R_{S}T}{1-b\rho} - \frac{a\alpha(T)\rho^2}{1+2b\rho-b^2\rho^2},\\
&\alpha(T) = [1+(0.37464+1.54226\omega - 0.26992\omega^2)(1-\sqrt{T/T_c})]^2
\end{split}
    \label{eq:SCpengrobinson}
\end{equation}
where $a = 0.45724R_{S}^2T_c^2/p_c$, $b = 0.0778R_{S}T_c/p_c$, $R_{S}$ is the specific gas constant, $p_c$, $T_c$ are the critical pressure and critical temperature of the pure substance, respectively, and $\omega$ is Pitzer's acentric factor of the substance. Note that the density $\rho$ in the Peng-Robinson EOS in Eq.~(\ref{eq:SCpengrobinson}) must be the density of a phase and not of a component. However, in the MCMP PP model by Bao and Schaefer~\cite{bao2013lattice}, the densities $\bar{\rho}_{\sigma}$ and $\bar{\rho}_{\bar{\sigma}}$ being substituted into EOSs to calculate $\psi_{\sigma}$ and $\psi_{\bar{\sigma}}$ are the cell-volume densities of individual components. These two concepts are different and we will detail their differences shortly. To distinguish the two types of densities, $\rho$ without an overbar is used to represent the density of a phase, or ``thermodynamic (bulk) density", while $\bar{\rho}$ with an overbar is used to represent the mass of an individual component within a grid cell volume, or ``cell-volume density". These notations are used through the whole paper.

\subsection{Thermodynamic inconsistencies in the available MCMP PP LB models}
The above MCMP PP LB models are only applicable for the cases with fully immiscible components. When applying such models to partially miscible MC mixtures, such as hydrocarbon fluids, there would be a number of pitfalls and the thermodynamic information behind phase behavior of fluids would not be correctly incorporated. These potential pitfalls are:
\begin{itemize}
    \item First, EOSs as written in Eq.~(\ref{eq:SCpengrobinson}) can describe the thermodynamic relationship between the pressure and the density only for a pure component, {\it i.e.}, when no other types of molecules are found in the system. When other molecules appear in the system, thermodynamic behavior of phases with multiple components will be different, and so would the EOS describing such behavior. This is particularly true when components have full or partial miscibility within co-existing phases, such as in the case for hydrocarbon mixtures. Strictly speaking, EOSs are always deployed to represent thermodynamic behavior of whole phases (liquid or vapor), and not of isolated components. 
    \item Second, it is important to realize that any density in an EOS represents the thermodynamic density, {\it i.e.}, ``mass of fluid phase per the actual volume occupied by that fluid phase". A higher thermodynamic density means a heavier phase. However, the density generated by LB models via summing up the local distribution functions is the cell-volume density, which means ``the mass of a component per the control volume of a grid cell". In a MC system, the cell-volume densities $\bar{\rho}_{\sigma} > \bar{\rho}_{\bar{\sigma}}$ at a spatial location just means there is more mass of component $\sigma$ than the mass of component $\bar{\sigma}$ inside the control volume of that location. These two density definitions are equivalent only for pure substances when the fluid phase consists only one component and occupies the entire grid cell. If a density is not the thermodynamic density, it should not be put into an EOS. As stated before, the concept of thermodynamic density in the context of EOSs applies exclusively to phases. In single phases where both component $\sigma$ and $\bar{\sigma}$ co-exist (due to miscibility), there is only one phase density and there are no separate thermodynamic densities for each component. Thermodynamic densities and cell-volume densities, can be numerically related through material balance statements.
    \item Third, at each spatial location, there is only one physically meaningful pressure. In SC PP models, the effective mass $\psi$ is usually defined by matching the thermodynamic pressure calculated from EOS and the hydrodynamic pressure derived in NSE~\cite{yuan2006equations}. However, in the MCMP PP model described above, there are three pressures: two EOS pressures $p_{EOS}(\bar{\rho}_{\sigma})$ and $p'_{EOS'}(\bar{\rho}_{\bar{\sigma}})$ (densities used to calculate the pressures in those models were cell-volume densities, which is inappropriate), plus the hydrodynamic pressure derived in NSE. Unless those three pressure definitions converge to the same value (and they obviously do not), the definition of effective mass using Eq.~(\ref{eq:effectivemass}) is not thermodynamically meaningful. 
\end{itemize}

\begin{figure}
\centerline
{\includegraphics[width=90mm]{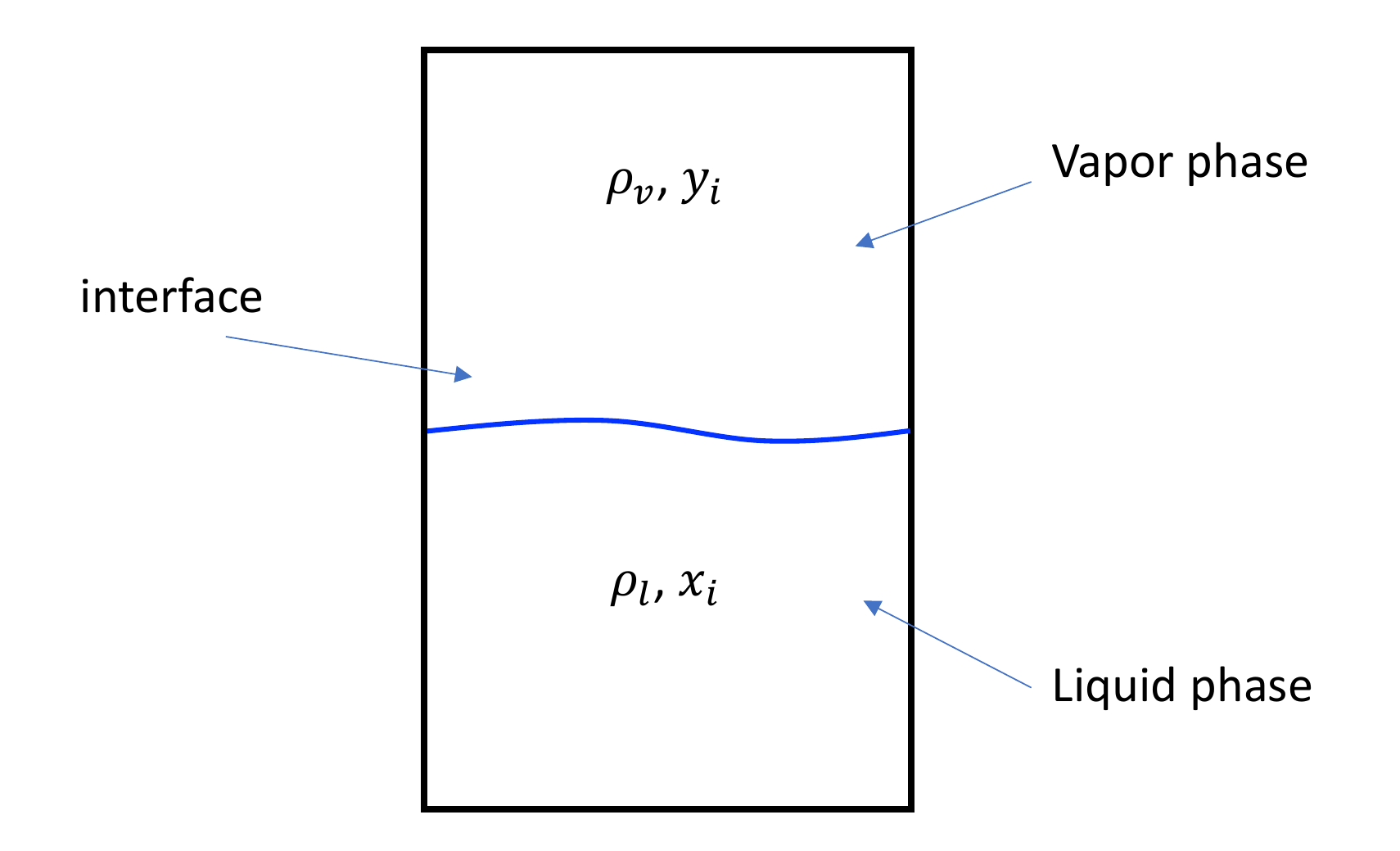}}
\vspace{0.1in}
\caption{A sketch of a liquid-vapor two-phase system.}
\label{fig:twophase}
\end{figure}

For illustration purposes, a multi-component multiphase system of partially miscible fluids of interest to the present study is sketched in Fig.~\ref{fig:twophase}. A vapor phase multi-component mixture and a liquid phase multi-component mixture are separated by an interface. Both vapor and liquid phases contain all components. This is why the system is referred as partially miscible, since within each phase, all components are fully miscible, but phases are immiscible. The vapor phase has component molar compositions of $y_{i}$, and the liquid phase has component molar compositions of $x_{i}$, $i = 1,2,\cdots,n_{c}$, $n_{c}$ is the total number of components in the system. The overall component molar composition of the whole system combining both phases is noted as $z_{i}$. The relationship between the vapor composition $y_{i}$, the liquid composition $x_{i}$, and the overall composition is
\begin{equation}
    z_{i} = y_{i}f_{ng} + x_{i}\left(1-f_{ng}\right),~~~i = 1,2,\cdots,n_{c}
\end{equation}
where $f_{ng}$ is the molar fractional of vapor that is found in the system. Assuming each phase is homogeneous, there is only one thermodynamically meaningful density (thermodynamic density) that is applied to each entire phase, {\it i.e.}, $\rho_{v}$ for the vapor phase and $\rho_{l}$ for the liquid phase. The cell-volume densities of components in each phase, while not so meaningful in a thermodynamic point of view, are related to the thermodynamic density of the phase as 
\begin{equation}
\begin{split}
    \bar{\rho}_{i,l} = \rho_{l}  \frac{x_{i}M_{i}}{\sum_{i}x_{i}M_{i}},~~~~\bar{\rho}_{i,v} = \rho_{v}\frac{y_{i}M_{i}}{\sum_{i}y_{i}M_{i}},\\
    \rho_{l}V = \sum_{i}\bar{\rho}_{i,l}V,~~~~\rho_{v}V = \sum_{i}\bar{\rho}_{i,v}V
    \end{split}
\end{equation} 
where $\bar{\rho}_{i,l}$ and $\bar{\rho}_{i,v}$ are the cell-volume densities of the $i$th component in the liquid and vapor phases, respectively, $M_{i}$ is the molar mass of component $i$, $V$ is the cell volume.

\section{A thermodynamically consistent MCMP PP model}\label{sec:model}
\subsection{A MCMP PP model}

As discussed earlier, an EOS should not be applied to individual components, but rather to phases, which are mixtures of miscible components.  Rather than relating the cell-volume density of each component to the thermodynamic pressure, the phase density of the MC mixture should be input to an EOS to compute thermodynamic pressure. Such implementation also avoids any potential discrepancy between thermodynamic density (phase density) and cell-volume density. In LB models, densities calculated from summing up distribution functions of each component are different from the density of the phase when more than one component exist at the same location. However, once all components are considered together as a whole, the two concepts of density are always equivalent. In addition, there is only one associated thermodynamic pressure (rather than one per component). Similar to the SCMP PP models, by matching this thermodynamic pressure with the hydrodynamic pressure derived in NSE, the effective mass in the PP models can be calculated. The third issue discussed earlier regarding multiple pressures at the same location can be also avoided. 

To achieve thermodynamic consistency, rather than using Eq.~(\ref{eq:MCMPforce}) to define the Shan-Chen force for each component, the total Shan-Chen force should be defined for the MC mixture as
\begin{equation}
        {\bf F}\left({\bf x}\right) = -G\psi\left({\bf x}\right)\sum_{\alpha}w_{\alpha}\psi\left({\bf x} + {\bf e}_{\alpha}\right){\bf e}_{\alpha},
        \label{eq:correctMCforce}
    \end{equation}
    where $G$ is the intensity of this force. Eq.~(\ref{eq:correctMCforce}) is actually the same definition of the Shan-Chen force in SCMP PP models. 
    With the total interaction force in Eq.~(\ref{eq:correctMCforce}), the hydrodynamic pressure derived in NSE would become~\cite{yuan2006equations} 
    \begin{equation}
      p_{{\rm hydro}} = c_s^2\rho + \frac{1}{2}G\delta t c_s^2\psi^2,
        \label{eq:hydropressure}
    \end{equation}
   where $\rho$ is the density of the phase consist of all components.  Eq.~(\ref{eq:hydropressure}) is also identical to its counterpart in SCMP PP models. Same as the SCMP PP models, the effective mass $\psi$ in Eq.~(\ref{eq:correctMCforce}) could be calculated by matching the hydrodynamic pressure in Eq.~(\ref{eq:hydropressure}) and thermodynamic pressure calculated from EOSs,
   \begin{equation}
      \psi = \sqrt{\frac{2\left(p_{EOS} - c_s^2\rho\right)}{G\delta t c_s^{2}}}.
        \label{eq:psicorrected}
    \end{equation}
    Per the discussion before, the thermodynamic pressure $p_{EOS}$ must be defined from an EOS applies to the MC system with the density of the phase, {\it i.e.}, input $\rho = \sum_{i}\bar{\rho}_{i}$.
    The interaction force applied to each mixture component (component 1, 2, 3, ..., $i$, ..., $n_c$) would just be a portion of this total interaction force, {\it i.e.}, 
    \begin{equation}
    \begin{split}
        &{\bf F}_{1}\left({\bf x}\right) = \chi_{1}{\bf F}\left({\bf x}\right),~~~{\bf F}_{2}\left({\bf x}\right) = \chi_{2}{\bf F}\left({\bf x}\right),~~~{\bf F}_{3}\left({\bf x}\right) = \chi_{3}{\bf F}\left({\bf x}\right)\cdots,\\
        &\sum_{i}\chi_{i} = 1,
         \end{split}
        \label{eq:forcedistribution}
    \end{equation}
    where $\chi_{i}$ is the force split factor of the $i$th component. 
    
    To the best of the authors' knowledge, the only previous attempt to split the total interaction force among mixture components has been presented by Gong {\it et al.}~\cite{gong2014lattice}. In the study of Gong {\it et al.}, it was reported that multiple empirical force split factors were attempted and the best candidate among them ended up being
    \begin{equation}
        \chi_{i} = \frac{c_{i}\sqrt{\hat{p}_{i}-p_{0}}}{\sum_{i}c_{i}\sqrt{\hat{p}_{i}-p_{0}}}
        \label{eq:Gong}
    \end{equation}
    where $c_{i}$ is the local molar composition, $\hat{p}_{i}$ and ${p}_{0}$ are the pressures computed from the non-ideal EOS and the ideal gas law, respectively, when the pure $i$th component has the same molar density as the mixture. There are a few potential problems with Gong {\it et al.}'s force split factor. First of all, as the non-ideal pressure is typically smaller than the ideal pressure due to molecular attraction, {\it i.e.}, $\hat{p}_{i}<p_{0}$, Eq.~(\ref{eq:Gong}) would lead to imaginary numbers. Second, there is no physical justification for such splitting choice, which ignores component volatility related to its interaction with other molecules. 
    
    When a MC fluid separates into two phases, components would have distinct preferences on which phase to accumulate in, {\it i.e.}, components have different volatility. For example, consider a two-component (binary) mixture of propane $C_3$ and pentane $nC_5$. Under a decreasing pressure at constant temperature below mixture's critical temperature, an initially liquid mixture starts to evaporate and form a vapor phase once it crosses its bubble point. In this process, heavier molecules of $nC_5$ will tend to remain in the liquid phase, while the lighter molecules of $C_3$ will have a stronger tendency to escape to the vapor. We describe $C_3$ as more volatile compared to $nC_5$, {\it i.e.}, it has preferential miscibility in the vapor phase. Molecules of volatile components typically have a smaller attractive or cohesive force (per unit mass) that prevents them from bonding tightly in a liquid form. Also, it is reasonable to expect that the force distributed to a specific component is proportional to the molar or mass fraction of this component locally. With these considerations, we propose the following force distribution strategy
\begin{equation}
\chi_{1} = \frac{\gamma_{1}\bar{\rho}_{1}V}{\sum_{i}\gamma_{i}\bar{\rho}_{i}V},~~~\chi_{2} = \frac{\gamma_{2}\bar{\rho}_{2}V}{\sum_{i}\gamma_{i}\bar{\rho}_{i}V},~~~\chi_{3} = \frac{\gamma_{3}\bar{\rho}_{3}V}{\sum_{i}\gamma_{i}\bar{\rho}_{i}V},\cdots,
    \label{eq:distributionforce}
\end{equation}
where $\gamma_{i}$ is the coefficient tied to the volatility of the $i$th component. For convenience, we set $\gamma_{i}$ of the least volatile (typically the heaviest) component to 1, which gives it the largest share of total attractive force relative to all other components. Thus $\gamma_{i} < 1$ for the rest of the components allocates a smaller share of the total attractive force per unit mass to more volatile components.

Now let us derive the mathematical expression for $\gamma_{i}$. Consider a two-component, two-phase fluid separated by a flat interface. At equilibrium, the hydrostatic balance is established as
\begin{equation}
    -\partial_{j}\left(c_s^2\rho\right) + F_{j} = 0,~~~~ F_{j}\approx -\partial_{j}\left(Gc_s^2\frac{\psi^2}{2}\right).
\end{equation}
At the same time, each component should also satisfy (assuming $\gamma_{2} = 1$)
\begin{subequations}
\begin{align}
   &- \partial_{j}\left(c_s^2\bar{\rho}_{1}\right) - \frac{\gamma_{1}\bar{\rho}_{1}}{\gamma_{1}\bar{\rho}_{1}+\bar{\rho}_{2}}\partial_{j}\left(Gc_{s}^{2}\frac{\psi^2}{2}\right) \approx -\partial_{j}\left(c_s^2\bar{\rho}_{1}\right)+\frac{\gamma_{1}\bar{\rho}_{1}}{\gamma_{1}\bar{\rho}_{1}+\bar{\rho}_{2}}F_{j}= 0,\\
    &- \partial_{j}\left(c_s^2\bar{\rho}_{2}\right) -  \frac{\bar{\rho}_{2}}{\gamma_{1}\bar{\rho}_{1}+\bar{\rho}_{2}}\partial_{j}\left(Gc_{s}^{2}\frac{\psi^2}{2}\right)\approx -\partial_{j}\left(c_s^2\bar{\rho}_{2}\right) +\frac{\bar{\rho}_{2}}{\gamma_{1}\bar{\rho}_{1}+\bar{\rho}_{2}}F_{j} = 0.
\end{align}
\label{eq:hydrostatic1}
\end{subequations}
The above two equations could also be rearranged as
\begin{subequations}
\begin{align}
    \frac{\gamma_{1}}{\gamma_{1}\bar{\rho}_{1}+\bar{\rho}_{2}}\partial_{j}\left(Gc_{s}^{2}\frac{\psi^2}{2}\right) + \partial_{j}\left(c_s^2\ln\bar{\rho}_{1}\right) \approx 0,\\
    \frac{1}{\gamma_{1}\bar{\rho}_{1}+\bar{\rho}_{2}}\partial_{j}\left(Gc_{s}^{2}\frac{\psi^2}{2}\right) + \partial_{j}\left(c_s^2\ln\bar{\rho}_{2}\right) \approx 0,
\end{align}
\label{eq:hydrostatic2}
\end{subequations}
which lead to
\begin{equation}
\gamma_{1}\partial_{j}\left(\ln \bar{\rho}_{2}\right) \approx \partial_{j}\left(\ln \bar{\rho}_{1}\right), 
\label{eq:hydrostatic3}
\end{equation}
Note that $\gamma_{1}$ is not spatially dependent. Therefore, integrating the above equation over the interface, we shall have
\begin{equation}
    \gamma_{1} \approx \frac{\ln\left(\bar{\rho}_{1,l}/\bar{\rho}_{1,v}\right)}{\ln\left(\bar{\rho}_{2,l}/\bar{\rho}_{2,v}\right)}  = \frac{\ln\left(\tilde{\rho}_{l}/\tilde{\rho}_{v}\right) - \ln{K_{1}}}{\ln\left(\tilde{\rho}_{l}/\tilde{\rho}_{v}\right) - \ln{K_{2}}},
    \label{eq:gamma}
\end{equation}
where $K_{1} = y_{1}/x_{1}$ and $K_{2} = y_{2}/x_{2}$ are the equilibrium (volatility) ratios of component 1 and component 2, respectively.
The value of $\gamma_{1}$ should be determined by the four cell-volume densities, or the molar densities ($\tilde{\rho}_{l}$ and $\tilde{\rho}_{v}$, with tilde caps in the notation) and molar compositions in both phases that satisfy thermodynamic equilibrium. 
According to the Gibbs phase rule, this information can be uniquely determined via the iso-fugacity rule when the temperature, pressure, and the overall composition are given, which will be covered shortly. As a result, $\gamma_{1}$ is also a function of the temperature, pressure, and the overall composition. This is somehow expected, since $\gamma_{i}$ is related to the volatility of components. Also, volatility is not constant, but must change with pressure, temperature, and composition, which is accounted for by the newly proposed splitting scheme.

    \subsection{Peng-Robinson EOS for multi-component hydrocarbon mixtures}
In order to correctly predict the phase behavior of MC hydrocarbon mixtures, it is also important to ensure that the EOS representing the correct MC thermodynamics is incorporated. 

In 1976, Peng and Robinson~\cite{peng1976new} summarized the following EOS for MC hydrocarbon system with $n_c$ components
\begin{equation}
    p_{EOS}\left(\tilde{\rho}\right) = \frac{\tilde{\rho} RT}{1-b_{m}\tilde{\rho}} - \frac{\left[a\alpha(T)\right]_{m}\tilde{\rho}^2}{1+2b_{m}\tilde{\rho}-b_{m}^2\tilde{\rho}^2},
    \label{eq:MCPengRobinson}
\end{equation}
where $\tilde{\rho}$ is the \emph{molar} density of the MC hydrocarbon mixture, $R$ is the universal gas constant. Parameters $\left[a\alpha\right]_{m}$, $b_{m}$ are defined for MC hydrocarbon mixtures using the following ``random mixing rules"~\cite{peng1976new}
\begin{equation}
    \begin{split}
        &\left[a\alpha\left(T\right)\right]_{m} = \sum_{i}^{n_c}\sum_{j}^{n_c}c_{i}c_{j}\sqrt{\left[a_{i}\alpha_{i}\left(T\right)\right]\left[a_{j}\alpha_{j}\left(T\right)\right]}\left(1-\zeta_{ij}\right),\\
        &b_{m} = \sum_{i}^{n_c}c_{i}b_{i},
    \end{split}
    \label{eq:randommixingrule}
\end{equation}
where $c_{i}$, $c_{j}$ are the molar composition of $i$th component and $j$th component in the mixture. 
When implementing EOS Eq.~(\ref{eq:MCPengRobinson}), $c_{i}$ and $c_{j}$ should be replaced by the local molar composition in a specific phase, {\it i.e.}, $x_{i}$ and $x_{j}$ in the liquid phase or $y_{i}$ and $y_{j}$ in the vapor phase, rather than the overall composition of the whole fluid. $\zeta_{ij}$ is the binary interaction coefficient between the $i$th and $j$th components, $\zeta_{ii} = \zeta_{jj} = 0$. For hydrocarbon components of similar nature, the interaction coefficient between any two components is also zero, $\zeta_{ij} = 0$~\cite{danesh1998pvt}. $a_{i}$, $\alpha_{i}\left(T\right)$, $b_{i}$ are the attraction parameter and the co-volume parameter for the $i$th component
\begin{equation}
    \begin{split}
        &a_{i} = \Omega_{ai}^{o}\frac{R^2T_{ci}^2}{p_{ci}},~~~~~b_{i} = \Omega_{bi}^{o}\frac{RT_{ci}}{p_{ci}},\\
        &\alpha_{i}\left(T\right) = \left[1+ m_{i}\left(1-\sqrt{T/T_{ci}}\right)\right]^2,
    \end{split}
\end{equation}
where $T_{ci}$, $p_{ci}$ are the critical temperature, and critical pressure of the $i$th component. $\Omega_{ai}^{o} = 0.45724$ and $\Omega_{bi}^{o} = 0.0778$ are constants~\cite{mccain2017properties,pedersen2006phase}. Finally, $m_{i}$ is defined through the Pitzer's acentric factor $\omega_{i}$ as~\cite{peng1976new}
\begin{equation}
    m_{i} =\left\{\begin{matrix}
&0.374640+1.54226\omega_{i}-0.26992\omega_{i}^{2},~~~~~~~~~~~~~~~~~~~~~~\omega_{i}\le0.49\\ 
&0.379642+1.48503\omega_{i}-0.164423\omega_{i}^{2}+0.016666\omega_{i}^{3},~~\omega_{i}>0.49
\end{matrix}\right. .
\end{equation}

The Peng-Robinson EOS that has been widely used in SCMP PP LB models ({\it i.e.}, Eq.~(\ref{eq:SCpengrobinson})) is just Eq.~(\ref{eq:MCPengRobinson}) applied to a pure substance. It is often convenient to recast Eq.~(\ref{eq:MCPengRobinson}) in terms of the compressibility factor $Z$, $Z = p/(\tilde{\rho} RT)$, as a cubic equation
\begin{equation}
Z^{3} + \left(B-1\right)Z^{2} + \left(A-2B^2-2B\right)Z -\left(AB-B^2-B\right) = 0,
    \label{eq:PRcubic}
\end{equation}
where 
\begin{equation}
\begin{split}
    &A = \sum_{i}^{n_c}\sum_{j}^{n_c}c_{i}c_{j}A_{ij},~~A_{ij} =\sqrt{A_{i}A_{j}}\left(1-\zeta_{ij}\right),\\&A_{i} =\Omega_{ai}^{o}\frac{p_{ri}}{T_{ri}}\left[1+ m_{i}\left(1-\sqrt{T_{ri}}\right)\right]^2,~~~~B = \sum_{i}^{i}c_{i}B_{i},~~~~B_{i} = \Omega_{bi}^{o}\frac{p_{ri}}{T_{ri}},
\end{split}
    \label{eq:cubiccoeff}
\end{equation}
where $p_{ri} = p/p_{ci}$, $T_{ri} = T/T_{ci}$ are the reduced pressure and reduced temperature of the $i$th component, respectively.
The cubic equation form of EOS in Eq.(\ref{eq:PRcubic}) is convenient to solve for $Z$, which will be used to calculate the fugacity and examine the thermodynamic consistency.

The thermodynamic consistency of SCMP PP models is usually examined by comparing the resulting liquid-vapor densities from a simulation with the benchmark results obtained from Maxwell equal area rule~\cite{he2002thermodynamic,shan2008pressure}. The Maxwell equal area rule is a statement of thermodynamic equilibrium between liquid and vapor phases when using cubic EOSs valid for pure substances~\cite{callen1998thermodynamics}. For a pure substance at any given temperature, the Maxwell equal area rule applies since the saturated vapor and saturated liquid must co-exist at the same pressure, which also corresponds to both bubble point and dew point pressures. However, for MC hydrocarbon mixtures, phase transitions occur over a range of pressures at any given temperature. For a mixture of known components at a given temperature, bubble point and dew point pressures are different, which makes the Maxwell equal area rule inapplicable. For MCMP PP model, the more general iso-fugacity criterion should be applied to determine whether the thermodynamic equilibrium is achieved. The Maxwell equal area rule is essentially a specific form of the iso-fugacity criterion applied to pure substances.

For a liquid-vapor two-phase system that is at thermodynamic equilibrium, the chemical potential of each component should be equal in all phases. The concept of fugacity is closely related to chemical potential, and can be used interchangeably for the purpose of computing phase equilibria~\cite{ertekin2018reservoir}. Fugacity is usually calculated via the fugacity coefficient as
\begin{equation}
    f_{vi} = \phi_{vi}y_{i}p,~~~~~~f_{li} = \phi_{li}x_{i}p,
    \label{eq:fugacity}
\end{equation}
where $f_{vi}$ and $f_{li}$ are the fugacity of the $i$th component in the vapor phase and liquid phase, respectively. $\phi_{vi}$ and $\phi_{li}$ are the corresponding fugacity coefficient, $p$ is the system pressure.

At a constant temperature, the fugacity coefficient of the $i$th component can be calculated in terms of the compressibility factor $Z$ as~\cite{peng1976new}
\begin{equation}
\begin{split}
    &\ln\phi_{i} = -\ln\left(Z-B\right) + \frac{B_{i}}{B}\left(Z-1\right)\\ &+\frac{A}{2\sqrt{2}B}\left(\frac{2\sum_{j}^{n_c}A_{ij}c_{j}}{A}-\frac{B_{i}}{B}\right)\ln\left[\frac{Z+\left(1-\sqrt{2}\right)B}{Z+\left(1+\sqrt{2}\right)B}\right].
    \end{split}
    \label{eq:fugacitycoeff2}
\end{equation}
The definitions of $A$, $B$, $A_{ij}$, and $B_{i}$ are given in Eq.~(\ref{eq:cubiccoeff}). When calculating the fugacity coefficients of $i$th component in the vapor and liquid phases, {\it i.e.}, $\phi_{vi}$ and $\phi_{li}$ via Eq.~(\ref{eq:fugacitycoeff2}), $c_{i}$ should be chosen as the molar composition in that particular phase, {\it i.e.}, $y_{i}$ for the vapor phase and $x_{i}$ for the liquid phase, instead of the overall molar composition combining the two phases. Similarly, the compressibility factor $Z$ must also be the unique value calculated for a specific phase. The value of $Z$ can be calculated from Eq.~(\ref{eq:PRcubic}) knowing temperature, pressure, and phase molar composition. When solving the cubic equation Eq.~(\ref{eq:PRcubic}), one can obtain either one or three real roots for $Z$. When there is only one real root of $Z$, this solution is used. When there are three real roots, the middle solution is always discarded. A typical approach is to select the smallest root of $Z$ (smallest compressibility) for the liquid phase and the largest root of $Z$ (largest compressibility) for the vapor phase, but the most reliable and thermodynamically consistent root selection criterion is to select the root of $Z$ which minimizes the associated Gibbs energy of the phase under consideration. This criterion can be mathematically expressed as~\cite{danesh1998pvt}:
\begin{equation}
\begin{split}
    &\frac{dG}{RT} = \left(Z_{max}-Z_{min}\right) + \ln\left(\frac{Z_{min}-B}{Z_{max}-B}\right) \\&+ \frac{A}{2\sqrt{2}B}\ln\left\{\frac{\left[Z_{min}+\left(1+\sqrt{2}\right)B\right]\left[Z_{max}+\left(1-\sqrt{2}\right)B\right]}{\left[Z_{min}+\left(1-\sqrt{2}\right)B\right]\left[Z_{max}+\left(1+\sqrt{2}\right)B\right]}\right\},
    \end{split}
    \label{eq:solutionselection}
\end{equation}
where $dG$ is the difference of Gibbs energy resulting from the largest solution of compressibility factor, $Z_{max}$, and the smallest solution $Z_{min}$. If $dG$ is positive, then $Z_{min}$ is selected, otherwise, $Z_{max}$ is selected. After the unique $Z$ corresponding to a specific phase is determined, the fugacity coefficient of each component in this phase is calculated via Eq.~(\ref{eq:fugacitycoeff2}), then the fugacity itself via Eq.~(\ref{eq:fugacity}). The fugacity of each component should be equal in the two phases if a thermodynamic equilibrium is achieved. The deviation from thermodynamic equilibrium can be quantified as a residual
\begin{equation}
    {\rm residual} = \sum_{i}^{n_{c}}\left(\frac{f_{li}}{f_{vi}} - 1\right)^2.
    \label{eq:error}
\end{equation}

The above iso-fugacity criterion is a more general way to examine the thermodynamic consistency of a MC mixture. It can also be used to guide the searching for the value of $\gamma_{1}$ in the present MCMP PP LB model. By running a MC phase behavior model at the given temperature, pressure and overall composition, molar densities and phase compositions (and thus the equilibrium ratios $K_{i}$) satisfying the iso-fugacity rule can be uniquely obtained.

\subsection{Thermodynamic inconsistency issue of the original Shan-Chen forces}

Although the proposed model introduces a correct way to incorporate the thermodynamic information of partially miscible multi-component fluids to the PP LB model, it may still not achieve full thermodynamic consistency in simulations. The remaining thermodynamic inconsistency comes from the combination of the Shan-Chen force in Eq.~(\ref{eq:correctMCforce}) and the definition of the effective mass $\psi$ using cubic EOSs in Eq.~(\ref{eq:psicorrected}), which has been widely recognized in SCMP PP models~\cite{kupershtokh2009equations,li2012forcing}. As the proposed MCMP PP model incorporates cubic EOS into the model in a similar way to those in SCMP PP models, this problem is inherited. Here we adopt the refined definition of Shan-Chen force proposed by Kupershtokh {\it et al.}~\cite{kupershtokh2009equations} to enforce thermodynamic consistency. Instead of using Eq.~(\ref{eq:correctMCforce}), the total Shan-Chen force is calculated as~\cite{kupershtokh2009equations,gong2012numerical}
    \begin{equation}
        {\bf F}\left({\bf x}\right) = -\beta G\psi\left({\bf x}\right)\sum_{\alpha}w_{\alpha}\psi\left({\bf x} + {\bf e}_{\alpha}\right){\bf e}_{\alpha} - \frac{1-\beta}{2}G\sum_{\alpha}w_{\alpha}\psi^2\left({\bf x} + {\bf e}_{\alpha}\right){\bf e}_{\alpha},
        \label{eq:coupledMCforce}
\end{equation}
where $\beta$ is a tuning parameter whose optimal value is found to achieve full thermodynamic inconsistency.

\subsection{The implementation of the proposed MCMP PP model}~\label{sec:implementation}

The implementation of the proposed MCMP PP model shares a lot in common with the SCMP PP models. 
\begin{itemize}
    \item With the local molar density of the phase and the local molar composition, the thermodynamic pressure $p_{EOS}$ is calculated with Eq.~(\ref{eq:MCPengRobinson}).
    \item By matching the obtained thermodynamic pressure $p_{EOS}$ with the hydrodynamic pressure defined in Eq.~(\ref{eq:hydropressure}), the effective mass $\psi$ is calculated with Eq.~(\ref{eq:psicorrected}).
    \item Then, the total Shan-Chen interaction force is calculated as Eq.~(\ref{eq:coupledMCforce}).
   \item After the total Shan-Chen force is defined, it is distributed to each component with the force split factor as Eq.~(\ref{eq:distributionforce}).
\end{itemize}
The determined interaction force for each component is then used to evolve the particle distribution functions of each component with LBE of that specific component. 

\section{Numerical validations and discussion}~\label{sec:validation}

\subsection{A limiting case}
A way to examine the capability of a MCMP PP LB model for multi-component multiphase partially miscible fluids is to benchmark it with a physically meaningful limiting case. When two components are essentially the same component, physical reality requires that a correct MCMP PP LB model should be able to reproduce the same results predicted by the SCMP PP LB model. For example, a ``mixture" of 50\% water and 50\% water must have precisely the same phase behavior as 100\% pure water under the same thermodynamic conditions. 

As a validation test, let us consider a simple case of a water droplet suspending in water vapor in a periodic domain. The thermodynamic behavior of water is captured by the reduced Peng-Robinson EOS for pure substances shown in Eq.~(\ref{eq:SCpengrobinson}). For demonstration purposes, we set the parameters $a = 2/49$, $b = 2/21$, and $R_{S} = 1$ for P-R EOS in the LB unit, as suggested by Yuan and Schaefer~\cite{yuan2006equations}. The acentric factor of water is $\omega = 0.344$. The computational domain has a size of $200 lu\times200 lu$, $lu$ stands for ``length unit". The relaxation time $\tau$ in the simulation is set to $1.0$. The droplet appears at the center of the domain, $(x_c,y_c) = (100,100)$, and has a radius $r_{0} = 30$. The initial phase density field follows 
\begin{equation}
  \rho\left(x,y,t=0\right) = \frac{\rho_l+\rho_v}{2}-\frac{\rho_{l}-\rho_{v}}{2}\tanh{\left[\frac{2\left(\sqrt{\left(x-x_{c}\right)^2+\left(y-y_{c}\right)^2}-r_{0}\right)}{W}\right]},
    \label{eq:initialden}
\end{equation}
where $\rho_{l}$ and $\rho_{v}$ are the saturated liquid and vapor phase densities at a given temperature $T$ below the critical temperature $T_{c}$, respectively. At a operation temperature $T = 0.9T_c$, $\rho_{l}$ and $\rho_{g}$ calculated from the Maxwell equal area rule are 5.90796 and 0.58007, respectively. $W = 8$ is the initial interface thickness, which is defined for initialization purposes and it might not maintain the same value as the simulation reaches its steady state. 

\begin{figure}
\centerline
{\includegraphics[width=200mm]{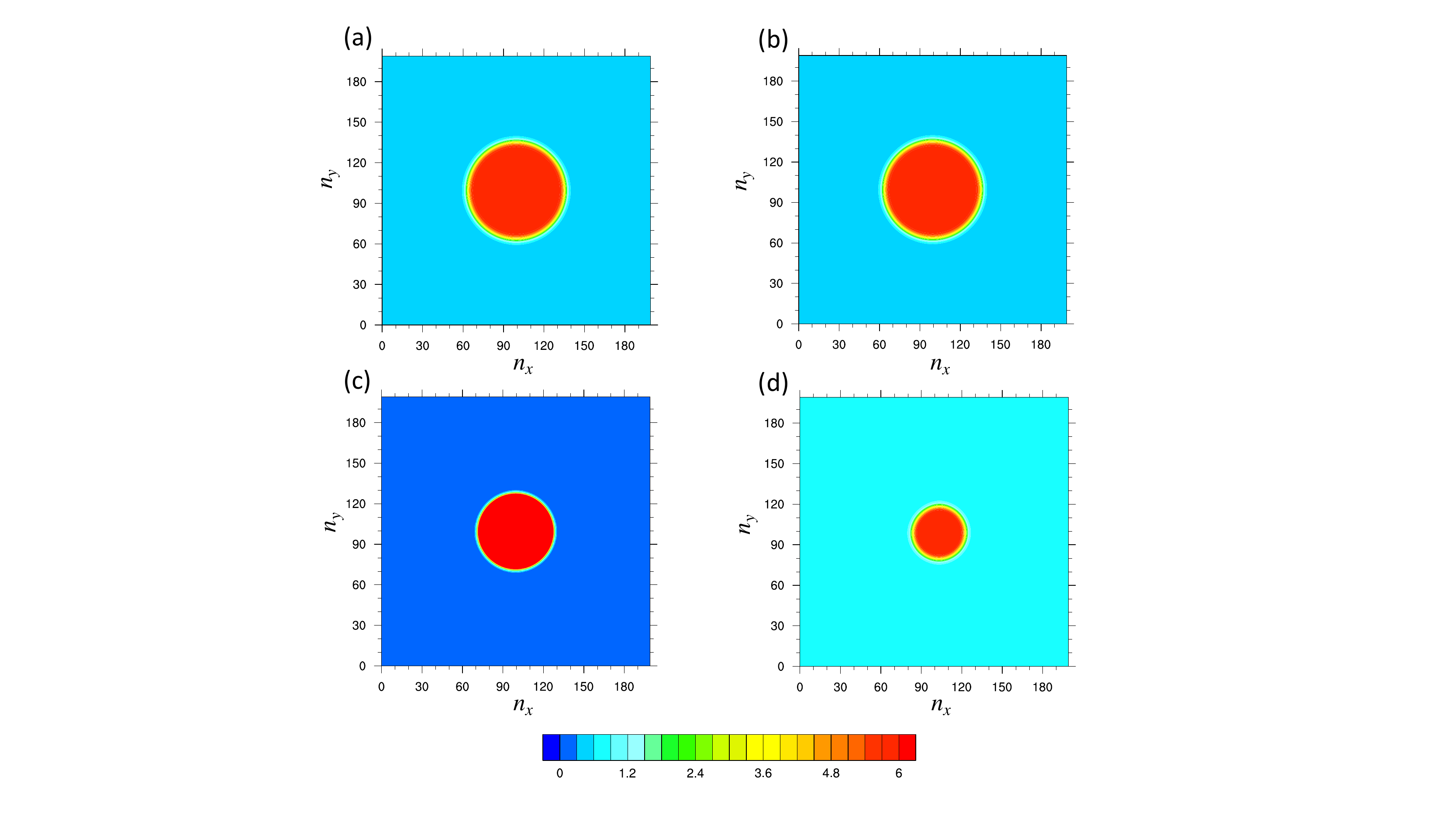}}
\vspace{0.1in}
\caption{Steady state density contours with SCMP model and MCMP model in the limiting case: (a) SCMP model, (b) proposed MCMP model, (c)  Bao and Schaefer~\cite{bao2013lattice}, $g_{\sigma\bar{\sigma}} = -0.05$, (d)  Bao and Schaefer~\cite{bao2013lattice}, $g_{\sigma\bar{\sigma}} = 0.05$.}
\label{fig:densitycontours}
\end{figure}

The above problem is simulated with both SCMP PP model and the proposed MCMP PP model with two identical components. In the simulation with the proposed MCMP PP model, for both component 1 and component 2, we use the same P-R EOS with $a_1 = a_2 = 2/49$, $b_1 = b_2 = 2/21$, $M_{1} = M_{2} = 1$, $R = 1$ in the LB unit, and $\omega = 0.344$. The relaxation time $\tau = 1.0$ is chosen to match the setting in the SCMP PP simulation. The initial cell-volume densities of component 1 and component 2 are set as 
\begin{equation}
\begin{split}
    \bar{\rho}_{1}\left(x,y,t=0\right) = \frac{\rho_{l}}{2} - \frac{\rho_{l}}{2}\tanh{\left[\frac{2\left(\sqrt{\left(x-x_{c}\right)^2+\left(y-y_{c}\right)^2}-r_{0}\right)}{W}\right]},\\
    \bar{\rho}_{2}\left(x,y,t=0\right) = \frac{\rho_{v}}{2} + \frac{\rho_{v}}{2}\tanh{\left[\frac{2\left(\sqrt{\left(x-x_{c}\right)^2+\left(y-y_{c}\right)^2}-r_{0}\right)}{W}\right]}.
\end{split}
    \label{eq:initialMC}
\end{equation}
Clearly, the summation of the two initial cell-volume densities matches precisely the initial phase density distribution in the SCMP model. Since the two components are essentially identical, they should have the same volatility, which leads to $\gamma_{1} = \gamma_{2} = 1$. At the steady state, the contours of the phase density $\rho$ ($\rho = \bar{\rho}_{1}+\bar{\rho}_{2}$ with the MCMP model) from the SCMP model and the proposed MCMP model are shown in Figure~\ref{fig:densitycontours}. As a comparison, results of simulations with Bao and Schaefer~\cite{bao2013lattice}'s MCMP model are also presented in parallel. In the SCMP model and the proposed MCMP model, the intensity of interaction force $G$ does not have an impact on the magnitude of the force, but is only required to ensure the term under the square root in Eq.~(\ref{eq:psicorrected}) is non-negative. On the other hand, in the MCMP model by Bao and Schaefer~\cite{bao2013lattice}, while the intensities of intra-molecular interaction forces $g_{\sigma\sigma}$ and $g_{\bar{\sigma}\bar{\sigma}}$ still have no influence on the magnitude of the intra-molecular interaction forces, the intensity of the inter-component interaction force $g_{\sigma\bar{\sigma}}$ does become relevant. In our tests, we set $g_{\sigma\sigma} = g_{\bar{\sigma}\bar{\sigma}} = -1$, and examine two values $g_{\sigma\bar{\sigma}} = -0.05$ and 0.05. In all these simulations, the equilibrium distributions of each component are defined with the cell-volume density of the component with Eq.~(\ref{eq:equilibrium}a), the forcing term in LBE is handled with Guo {\it et al.}'s forcing scheme~\cite{guo2002discrete} in Eq.~(\ref{eq:equilibrium}b).

\begin{figure}
\centerline
{\includegraphics[width=80mm]{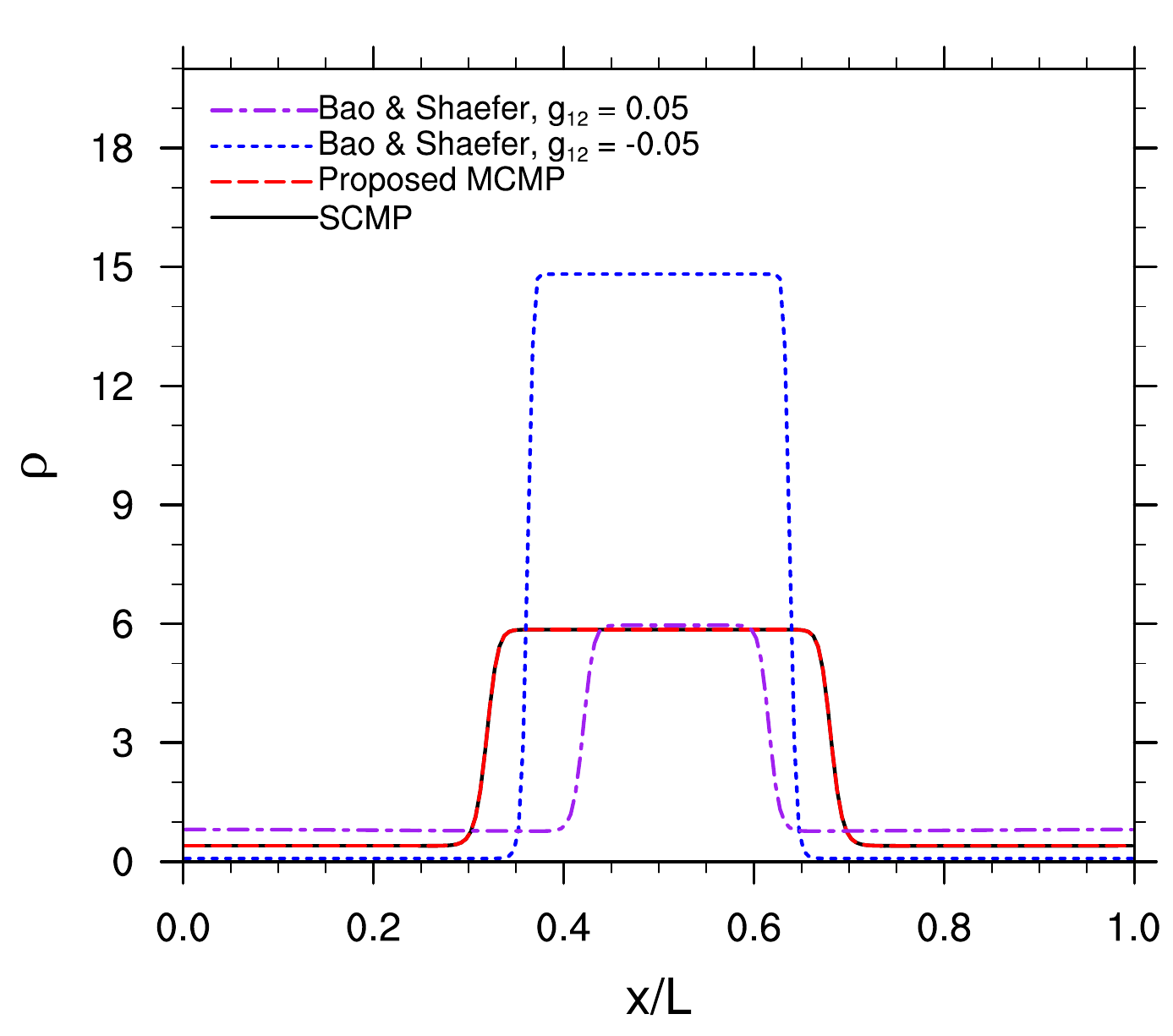}}
\vspace{0.1in}
\caption{Steady state density distribution on a line cross the droplet center. $g_{12}$ in the figure legend means $g_{\sigma\bar{\sigma}}$ in the text in Bao and Schaefer's MCMP model.}
\label{fig:densityline}
\end{figure}

As shown in Figure~\ref{fig:densitycontours}, only the proposed MCMP model results in identical density distribution as the SCMP model in this limiting case. The results from Bao and Schaefer's MCMP model~\cite{bao2013lattice}, however, significantly deviate from the phase density predicted by the SCMP model. The distributions of phase density on a line cutting through the center of the droplet are shown in Fig.~\ref{fig:densityline}, which further confirm such deviations. While this limiting case is simple, it clearly indicates that the MCMP model proposed by Bao and Schaefer~\cite{bao2013lattice}, as well as those models developed on similar bases~\cite{yu2011numerical,kamali2013simulating,chen2015pore}, are not able to predict the phase behavior of
partially miscible multi-component fluid due to their false ways to incorporate thermodynamic information. Thus, the use of these MCMP models for multi-component, multiphase partially miscible mixtures is not advisable. However, it is expected that these models would remain useful to examine behaviors of immiscible systems where components would not mix and would remain fully separated in different phases, but the potential misuse of EOSs in those models still requires further attention.

\subsection{Two-component, two-phase hydrocarbon mixture separated by a flat interface}

\begin{figure}
\centerline
{\includegraphics[width=80mm]{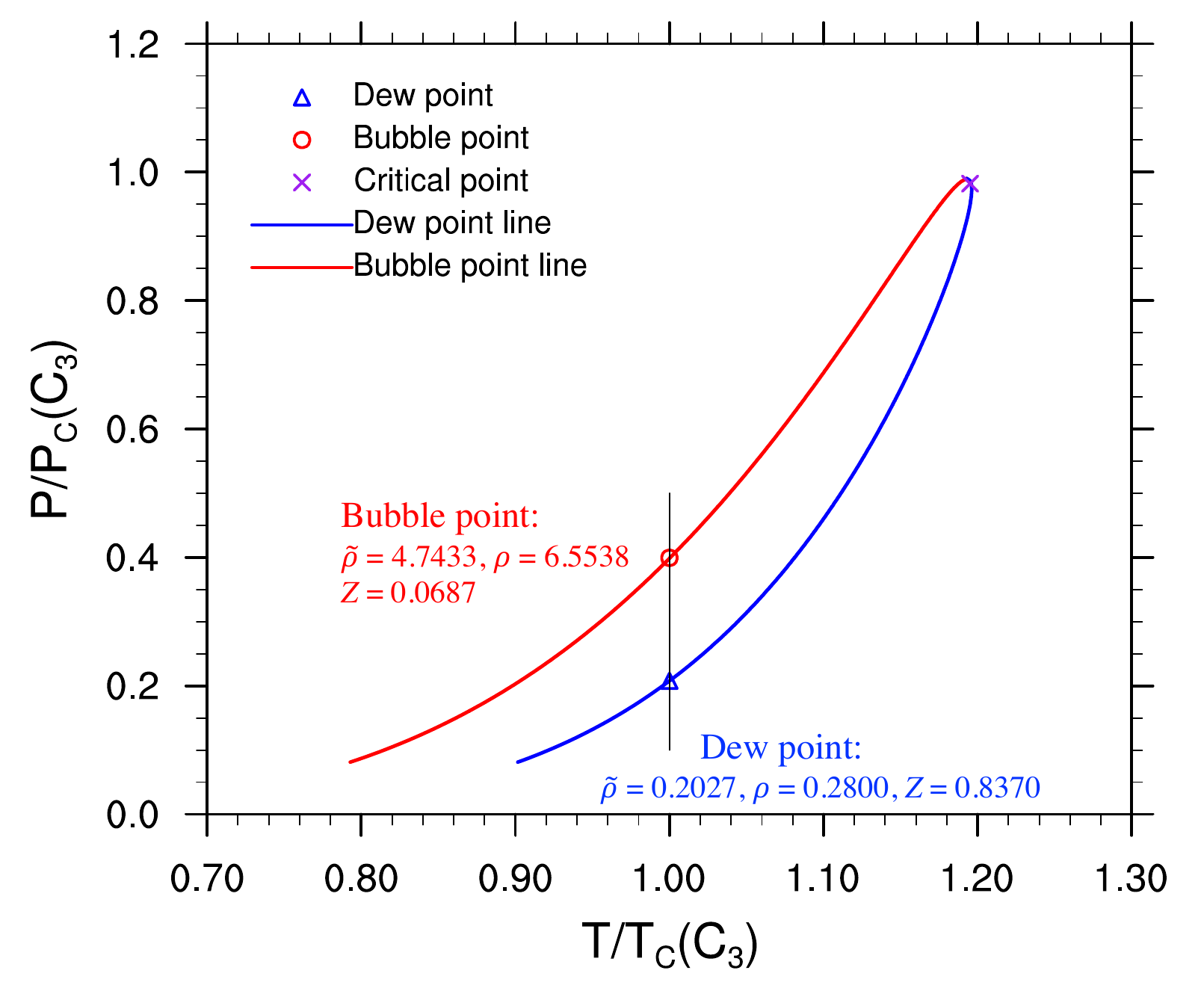}}
\vspace{0.1in}
\caption{A phase diagram of two-component mixture of 40\% propane $C_3$ (component 1) and 60\% pentane $nC_5$ (component 2). The densities at the bubble and dew point in the figure have been scaled to the LB units using $a_{1} = 2/49$, $b_{1} = 2/21$, $M_{1} = 1$, and $R = 1$.}
\label{fig:phaseenvelope}
\end{figure}

We now test the proposed MCMP PP model with a realistic two-component hydrocarbon mixture. Unlike pure substances, whose phase transition happens only at a specific pressure at a given temperature, the phase transition of a two-component hydrocarbon mixture of a given molar composition occurs within a range of pressures at a given temperature. This is shown in the pressure-temperature (PT) diagram of this multi-component mixture via a ``phase envelope". A phase envelope is a curve that encloses the temperatures and pressures under which the mixture of a given molar composition will separate into two phases. As shown in Fig~\ref{fig:phaseenvelope}, the red solid curve is the bubble point line, whose left-side indicates the pressure-temperature conditions under which the hydrocarbon mixture remains single-phase liquid. The blue solid curve is the dew point line, whose right-side indicates the pressure-temperature conditions under which the mixture remains single-phase vapor. The intersection of the bubble point line and the dew point line is the critical point, at which the liquid and vapor phases are indistinguishable. It should be noted that the phase envelope shown in Fig.~\ref{fig:phaseenvelope} does not consider the presence of capillary pressure introduced by curved interfaces. When capillary pressure is present, phase-envelope locations will change, especially at conditions away from the critical point~\cite{nojabaei2013effect,sandoval2015phase}.

\begin{table*}
\caption{Thermodynamic properties of hydrocarbon components, from first column: component, critical pressure, critical temperature, Pitzer's acentric factor, molar mass, specific gas constant~\cite{danesh1998pvt}. }
\footnotesize
\begin{center}
\begin{tabular}
{c c c c c c c c c c c}
\hline
\toprule
& \begin{tabular}{@{}c@{}}$p_{ci}$ \\ (psia)\end{tabular}&\begin{tabular}{@{}c@{}}$T_{ci}$\\$(^{\circ} R)$\end{tabular}& $\omega_{i}$&\begin{tabular}{@{}c@{}}$M_{i}$\\(lbm/lbmol)\end{tabular}&\begin{tabular}{@{}c@{}}$R_{i}$\\(psia ft$^3$/lbm $^{\circ} R$)\end{tabular}\\
\midrule 
$C_1$& 666.40&343.33&0.0104&16.043&0.669\\
$C_2$& 706.50&549.92&0.0979&30.070&0.357\\
$C_3$& 616.00&666.06&0.1522&44.097&0.243\\
$iC_4$& 527.90&734.46&0.1822&58.123&0.185\\
$nC_4$& 550.60&765.62&0.1995&58.123&0.185\\
$iC_5$& 490.40&829.10&0.2280&72.150&0.149\\
$nC_5$& 488.60&845.80&0.2514&72.150&0.149\\
$nC_6$& 436.90&913.60&0.2994&86.177&0.125\\
$C_7+$& 305.20&1112.00&0.4898&142.285&0.075\\
\bottomrule
\end{tabular} 
\end{center}
\label{tab:thermodynamicproperties}
\end{table*} 

\begin{table*}
\caption{Thermodynamic properties and the resulting force splitting coefficients of a 40\% $C_3$ (component 1) and 60\% $nC_5$ (component 2) mixture at $T = T_{c,C_3}$ under different pressures.}
\footnotesize
\begin{center}
\begin{tabular}
{c c c c c c c c c c c c c c c}
\hline
\toprule
\begin{tabular}{@{}c@{}}$p$ \\ (psia)\end{tabular}&$Z_{l}$&$Z_{v}$&$x_{1}$&$y_{1}$&\begin{tabular}{@{}c@{}}$\bar{\rho}_{1,l}$\\(lb/ft$^3$)\end{tabular}&\begin{tabular}{@{}c@{}}$\bar{\rho}_{1,v}$\\(lb/ft$^3$)\end{tabular}&\begin{tabular}{@{}c@{}}$\bar{\rho}_{2,l}$\\(lb/ft$^3$)\end{tabular}&\begin{tabular}{@{}c@{}}$\bar{\rho}_{2,v}$\\(lb/ft$^3$)\end{tabular}&\begin{tabular}{@{}c@{}}$\gamma_{1}$\\($\gamma_{2} = 1$)\end{tabular}\\
\midrule 
246.0& 0.06877&(-)&0.40000&(-)&8.829&(-)&21.668&(-)&(-)\\
240.0& 0.06716&0.76441&0.38792&0.71917&8.554&22.083&1.393&0.890&0.56515\\
220.0& 0.06180&0.77837&0.34334&0.68507&7.542&23.601&1.195&0.899&0.56378\\
200.0& 0.05644&0.79194&0.29768&0.64462&6.508&25.124&1.004&0.906&0.56242\\
180.0& 0.05108&0.80509&0.25094&0.59567&5.457&26.650&0.822&0.913&0.56107\\
160.0& 0.04568&0.81779&0.20313&0.53506&4.390&28.175&0.646&0.918&0.55973\\
140.0& 0.04025&0.82993&0.15422&0.45783&3.310&29.699&0.477&0.923& 0.55839\\
127.9&(-)&0.83699&(-)&0.40000&(-)&(-)&0.377&0.926&(-)\\
\bottomrule
\end{tabular} 
\end{center}
\label{tab:fugacity}
\end{table*} 

\begin{figure}
\centerline
{\includegraphics[width=140mm]{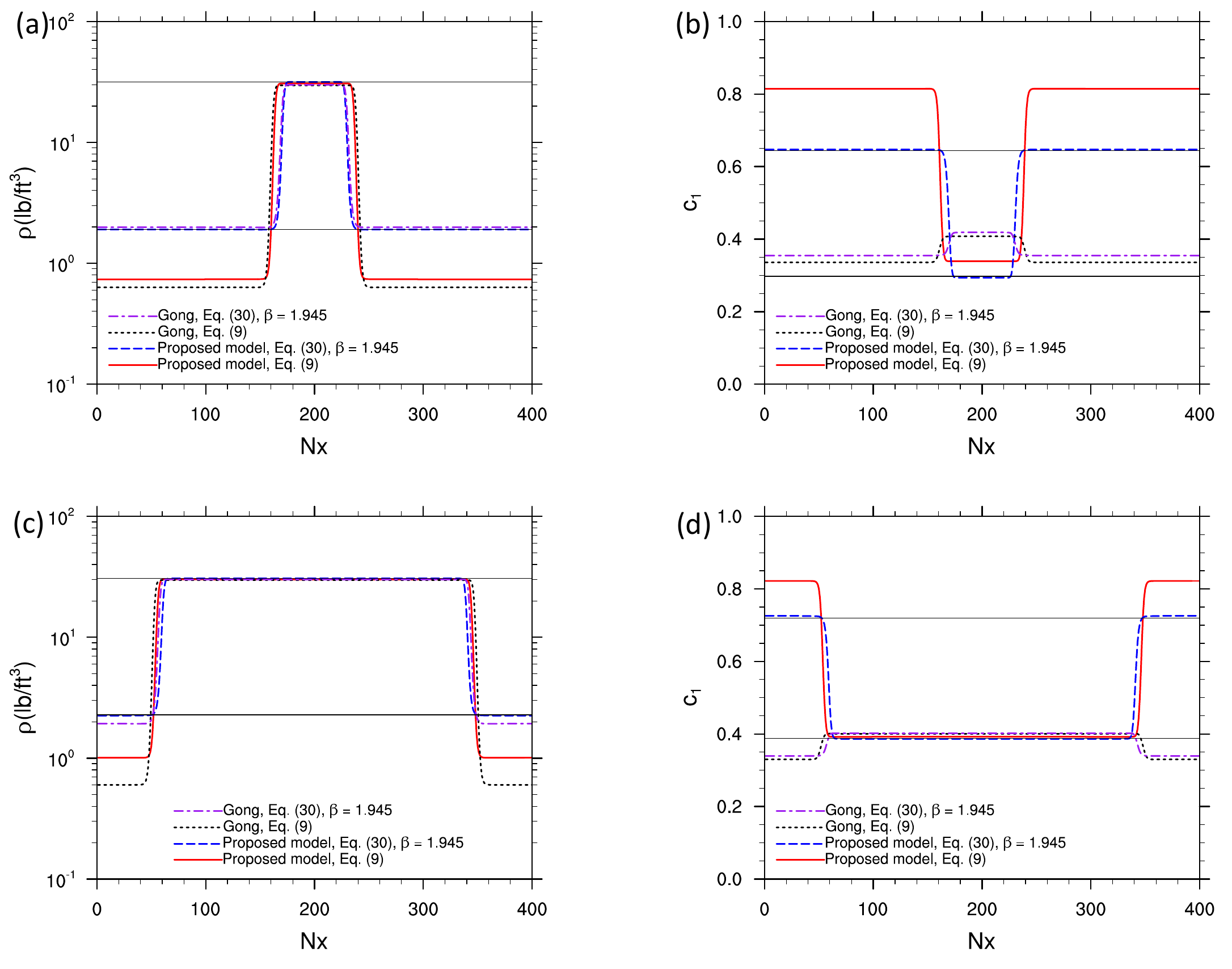}}
\caption{Density and molar distribution of a 40\% propane and 60\% pentane mixture at $T = T_{c,C_{3}}$ under two pressures $p = 200$ psia and 240 psia. Subfigures: top row: $p = 200$ psia; bottom row: $p = 240$ psia; left column: phase density distribution; right column: molar composition distribution. The corresponding results satisfying thermodynamic consistency are represented by the horizontal thin solid lines in each subfigure.}
\label{fig:correctedthermo}
\end{figure}

In this case study, we simulate a two-component, two-phase fluid separated by a flat interface. The first component is propane $C_{3}$ and the second component is pentane $nC_{5}$. For readers' convenience, the thermodynamic properties of the common hydrocarbon components are listed in Table~\ref{tab:thermodynamicproperties}. The overall molar composition of the mixture is $z_{1} = 0.4$ and  $z_{2} = 0.6$. The prevailing temperature is set to $T = T_{c,C_{3}} = 666.06 ^{\circ}R$. With the given temperature and overall composition, the cell-volume densities that satisfy iso-fugacity rule and the resulting $\gamma_{1}$ under different pressures are given in Table~\ref{tab:fugacity}. For this specific setup, $\gamma_{1}$ is close to a constant under the pressure range that would lead to phase separation. With the values of $\gamma_{1}$ obtained from the thermodynamic properties of the hydrocarbon mixture, the performance of the proposed MCMP PP model is tested at two selected pressures $p = 200~{\rm psia}$ and $p = 240~{\rm psia}$. The simulations are conducted in a periodic domain of $nx\times ny = 400\times 2$, and the initial cell-volume density distribution of each component is set as
\begin{equation}
    \bar{\rho}_{i}\left(x,y,t = 0\right) =  \left\{\begin{matrix} 
    &\dfrac{\bar{\rho}_{i,l}+\bar{\rho}_{i,v}}{2} + \dfrac{\bar{\rho}_{i,l}-\bar{\rho}_{i,v}}{2}\tanh{\left[\dfrac{2\left(x-L_{0}\right)}{D}\right]},~~~~x\le L/2\\
    \\
    &\dfrac{\bar{\rho}_{i,l}+\bar{\rho}_{i,v}}{2} - \dfrac{\bar{\rho}_{i,l}-\bar{\rho}_{i,v}}{2}\tanh{\left[\dfrac{2\left(x-L+L_{0}\right)}{D}\right]},~~~~x > L/2 \end{matrix}\right .
    \label{eq:initialdensity}
\end{equation}
where $D = 8$, $L = 400$, and
\begin{equation}
    L_{0} = \frac{\left(\bar{\rho}_{1,l}z_{1}+\theta\bar{\rho}_{2,l}z_{1}-\bar{\rho}_{1,l}\right)L}{2\left(\bar{\rho}_{1,v}-\bar{\rho}_{1,l}-\bar{\rho}_{1,v}z_{1}-\theta\bar{\rho}_{2,v}z_{1}+\bar{\rho}_{1,l}z_{1}+\theta\bar{\rho}_{2,l}z_{1}\right)},
    \label{eq:initialdensity2}
\end{equation}
$\theta = M_{1}/M_{2}$ is the ratio of molecular weight. The cell-volume densities $\bar{\rho}_{i,l}$, $\bar{\rho}_{i,v}$, $i = 1,2$ are given in Table~\ref{tab:fugacity} under the given pressures. Eq.~(\ref{eq:initialdensity2}) is used to make sure the desired overall composition $z_{1}$ is obtained with the initial cell-volume density distribution. The parameters in the LB simulations are still scaled from the physical scaling with $a_{1} = 2/49$, $b_{1} = 2/21$, $M_{1} = 1$, and $R = 1$. Then the values of $a_{2}$, $b_{2}$, $M_{2}$ are defined accordingly. This setup is also adopted in all later validation test cases. The relaxation time $\tau$ is set to $0.8$. The equilibrium distributions and the forcing term in LBE are defined by the corresponding expression in Eq.~(\ref{eq:equilibrium}). At the steady state, the resulting density and composition distribution are presented in Fig.~\ref{fig:correctedthermo}. Here we compare the results from four different realizations, 1) the proposed MCMP PP model with the total Shan-Chen force defined by Eq.~(\ref{eq:correctMCforce}), 2) the proposed MCMP PP model with the Shan-Chen force defined by Eq.~(\ref{eq:coupledMCforce}) with $\beta = 1.945$, 3) Gong {\it et al.}'s MCMP PP model~\cite{gong2014lattice} with the total Shan-Chen force defined by Eq.~(\ref{eq:correctMCforce}), and 4) Gong {\it et al.}'s MCMP PP model with the Shan-Chen force defined by Eq.~(\ref{eq:coupledMCforce}) with $\beta = 1.945$.  It is worth emphasizing that for Gong {\it et al}.'s force split factor to be calculated, the two pressures under the square root in Eq.~(\ref{eq:Gong})  must be swapped to make the whole term positive. 

As shown in Figure~\ref{fig:correctedthermo}, with Gong {\it et al}.'s force split factor, the predicted molar composition distribution is quantitatively and qualitatively incorrect as the light component $C_{3}$ has higher molar fraction in the liquid phase compared to the vapor phase. On the contrary, the proposed force split factor always ensures that the light component would have a higher molar fraction in the vapor phase. When the total Shan-Chen force is calculated as its original definition in Eq.~(\ref{eq:correctMCforce}), certain thermodynamic inconsistency is still observed. This remaining thermodynamic inconsistency can be further suppressed via the modified definition of the Shan-Chen force in Eq.~(\ref{eq:coupledMCforce}) with a proper choice of the tuning parameter $\beta = 1.945$. It should be noted that this optimal value of $\beta$ is specifically obtained with Guo et al.'s~\cite{guo2002discrete} forcing scheme in LBM. It may change if another forcing scheme is adopted. 

\begin{figure}
\centerline
{\includegraphics[width=140mm]{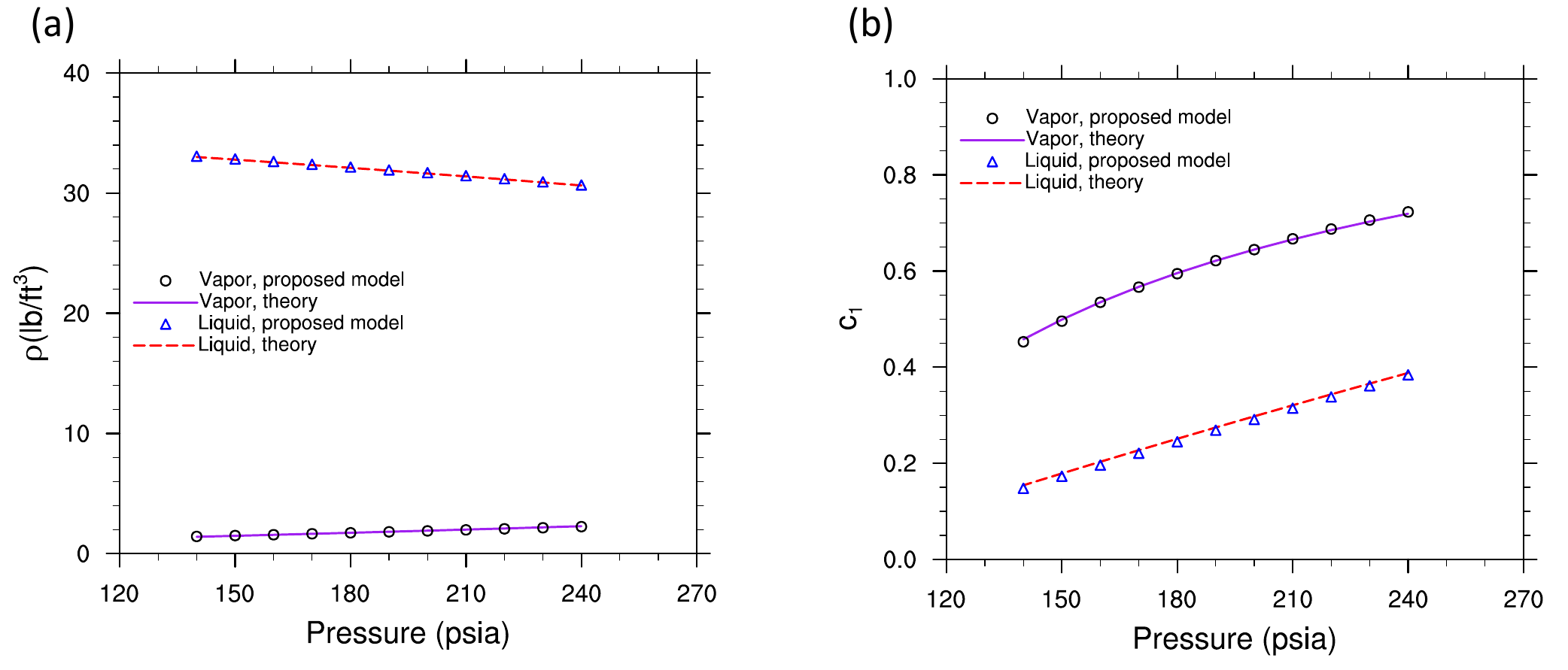}}
\caption{Comparisons of the (a) phase density, and (b) molar composition between the LB simulation with the physical values satisfying the iso-fugacity rule.}
\label{fig:comparisonthermo}
\end{figure}

Next, the resulting phase density and the molar composition in each phase under different pressures for the 40\% $C_{3}$ and 60\% $nC_{5}$ mixture at $T = T_{c,C_3}$ are compared with the corresponding physical value in Figure~\ref{fig:comparisonthermo}. These simulations are conducted with the identical setting as before expect that the computational domain is extended to $nx\times ny = 1600\times 2$ to create a larger region of the liquid phase under small pressures. The total Shan-Chen forces in all these simulations are again defined with Eq.~(\ref{eq:coupledMCforce}) with $\beta = 1.945$. It can be clearly seen that the LB simulation results match quite well with the benchmark results satisfying the iso-fugacity rule. The maximum relative error for the molar composition is $4.0\%$ and the maximum relative error for the phase density is $2.1\%$. 
This comparison indicates that the present MCMP PP LB model is able to honor the thermodynamics of MC hydrocarbon mixtures.   

\begin{figure}
\centerline
{\includegraphics[width=90mm]{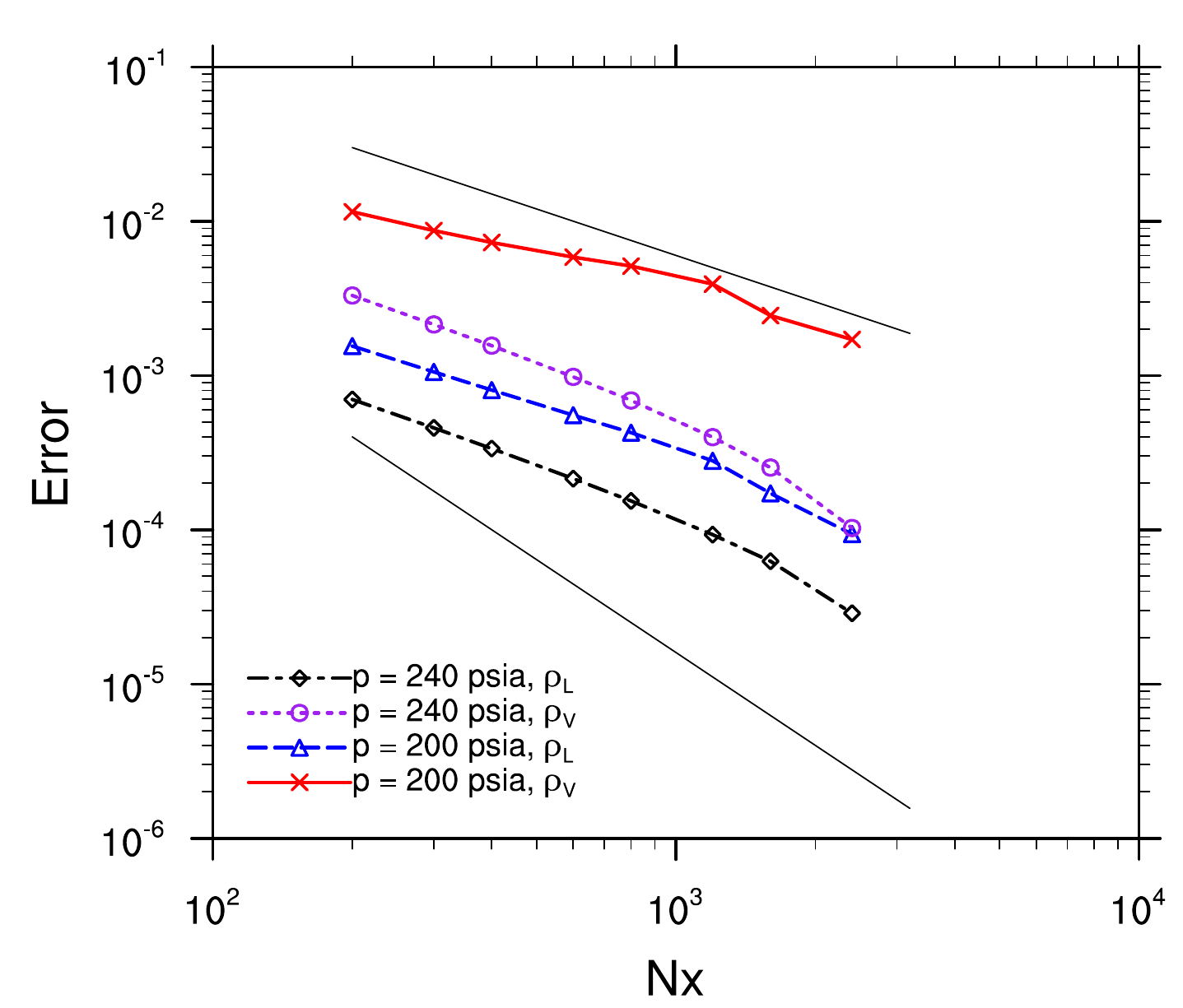}}
\caption{Errors of phase densities in the case of a two-component, two-phase hydrocarbon fluid as functions of grid size. The two thin straight lines are the reference lines for the first and second-order convergence rate.}
\label{fig:convergencerate}
\end{figure}

This validation case is further utilized to conduct a spatial convergence study on the simulation results of phase density. For this 40\% $C_3$ and 60\% $nC_5$ mixture at $T = T_{c,C_3}$, we simulate the same flat interface case with different mesh sizes from $nx = 200$ to $nx = 3200$, again at two selected pressures $p = 200$ and $240$ psia. The relative errors of phase densities compared to the results with the highest grid resolution, {\it i.e.}, $|\left(\rho_{nx} - \rho_{3200}\right)/\rho_{3200}|$ are plotted in Figure.~\ref{fig:convergencerate}. The spatial accuracy of the phase density has a slight dependency on the specific thermodynamic conditions chosen for the simulation, but being roughly first-order.

\subsection{Two-component, two-phase hydrocarbon mixture separated by curved interfaces}
We now move to more general cases with curved interface. The third validation test is a stationary droplet suspending in the vapor phase. Here we mainly validate the case against the Laplace's law, which for a two-dimensional droplet states $\Delta p = \sigma/R$, where $\Delta p = p_l - p_v$ is the pressure difference between the liquid droplet and its vapor environment, $\sigma$ is the surface tension, and $R$ is the radius of the liquid droplet. For a given two-phase system of multi-component hydrocarbon fluids at thermodynamic equilibrium, the surface tension should remain a constant, which does not change with the radius of the droplet. According to the Laplace law, this would require the pressure difference $\Delta p$ to decrease inverse proportionally as the droplet radius increases.

The simulations are conducted in different sizes of periodic computational domains to accommodate different size of droplet. The size of the computational domain varies from $nx\times ny = 50\times 50$ to $800\times 800$, and for each size, the droplet has an initial radius of $R_0 = nx/5$ and locates at the center of domain. The initial distribution of the cell-volume density is set to
\begin{equation}
 \bar{\rho}_{i}\left(x,y,t = 0\right) =  \dfrac{\bar{\rho}_{i,l}+\bar{\rho}_{i,v}}{2} + \dfrac{\bar{\rho}_{i,l}-\bar{\rho}_{i,v}}{2}\tanh{\left[\dfrac{2\left(\sqrt{\left(x-x_c\right)^2+\left(y - y_c\right)^2}-R_{0}\right)}{W}\right]},
    \label{eq:initialdensity_droplet}   
\end{equation}
where the ${\bar\rho}_{i,l}$ and ${\bar\rho}_{i,v}$, $i = 1,2$ are chosen from Table~\ref{tab:fugacity} for $p = 200$ and $240$ psia, $(x_c,y_c)$ is the center location of the droplet, and the initial width of the interface $W$ is set to 4. For this case with curved interface, $\beta$ in Eq.~(\ref{eq:coupledMCforce}) is slightly adjust to 1.915 from its optimal value 1.945 in the flat interface case to ensure thermodynamic consistency. The relaxation time $\tau = 0.8$ is still applied in the simulations.

\begin{figure}
\centerline
{\includegraphics[width=90mm]{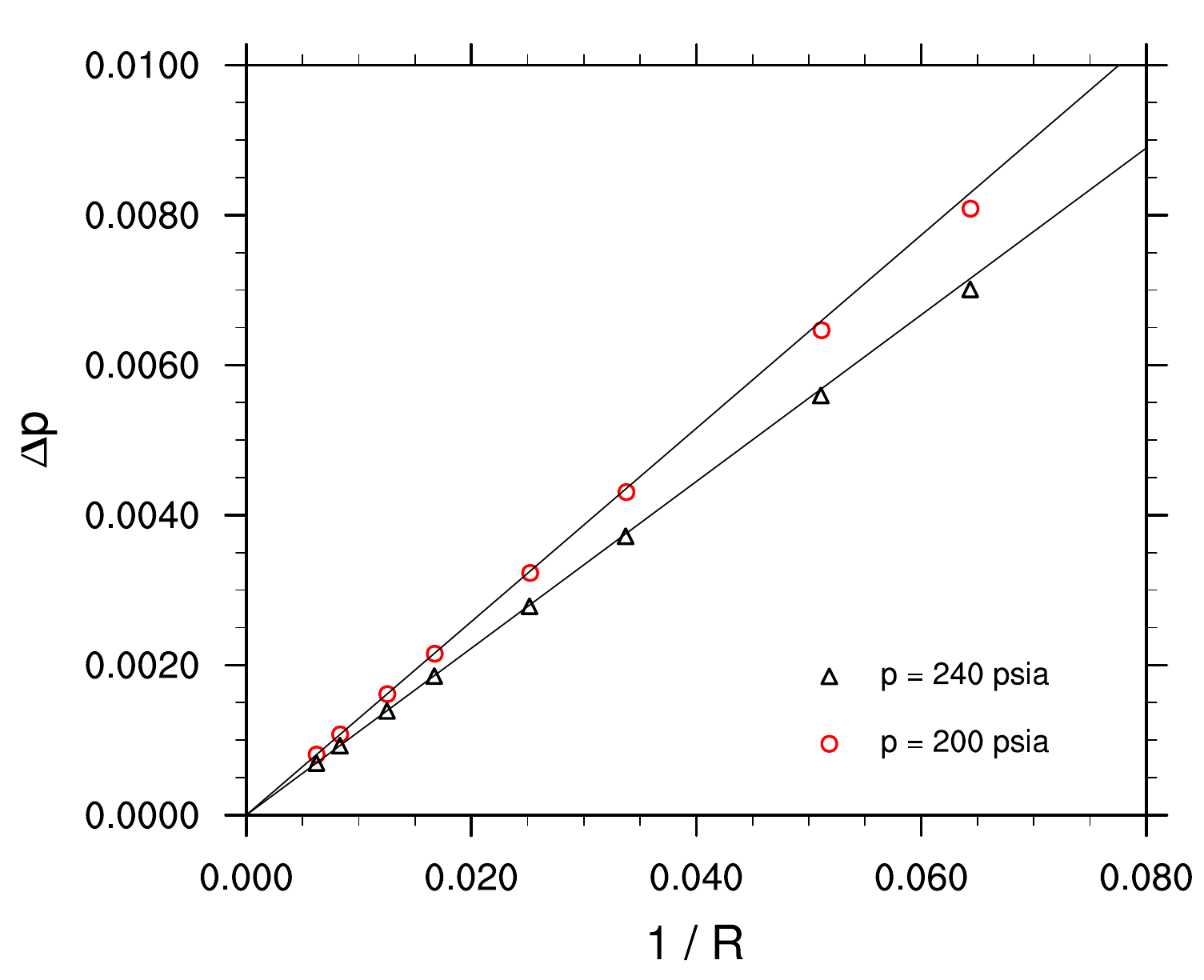}}
\caption{The validation of the Laplace's law with the proposed MCMP PP LB model.}
\label{fig:surfacetension}
\end{figure}

\begin{figure}
\centerline
{\includegraphics[width=90mm]{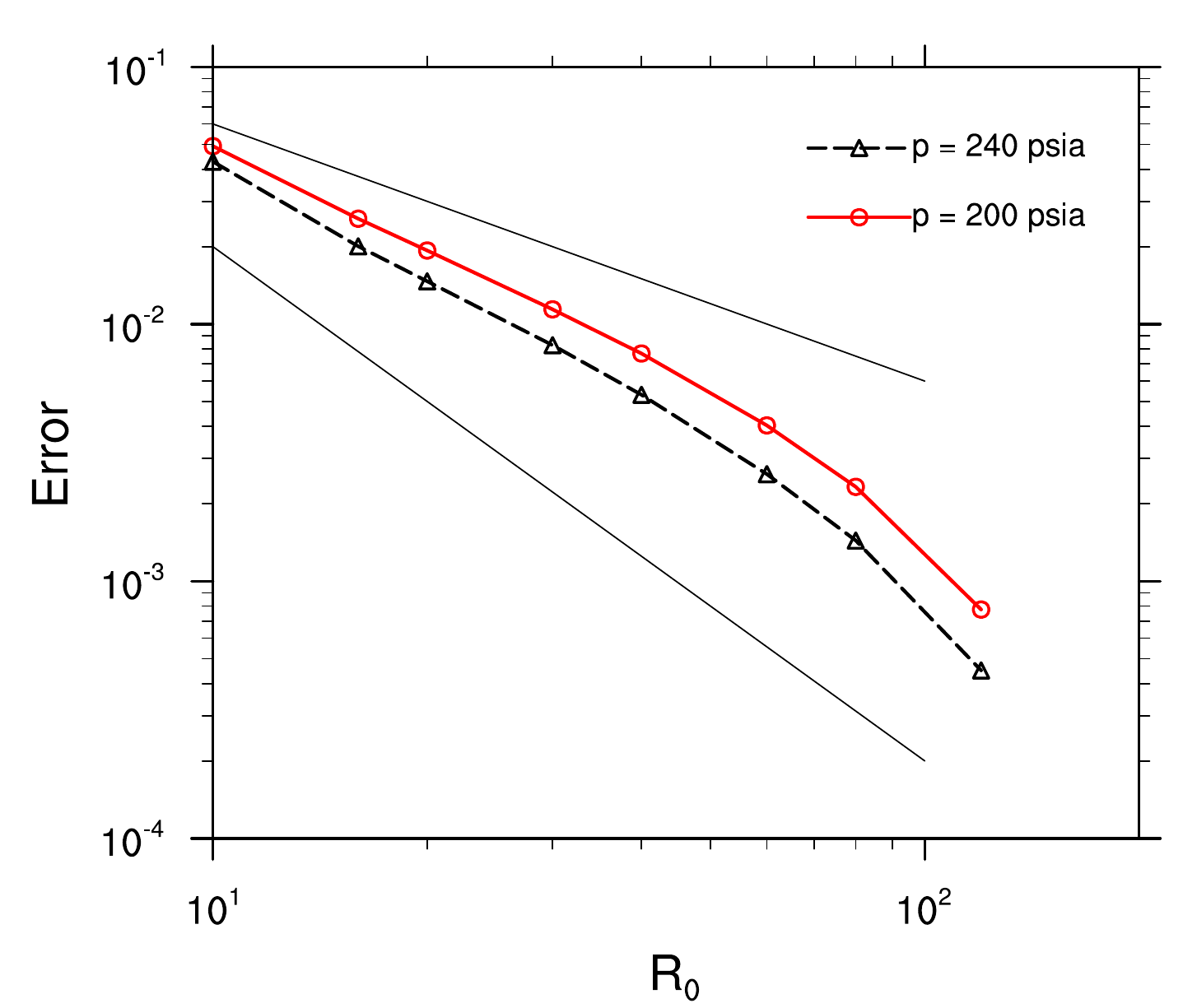}}
\caption{The spatial convergence rate of the predicted surface tension through Laplace's law. The two thin straight lines are the reference lines for the first and second-order convergence rate.}
\label{fig:surfacetensionconvergencerate}
\end{figure}

At the steady state, the radii of droplet, which are measured by the location of $0.5(\rho_l + \rho_v)$ in each case, and the pressure differences are recorded. The relationships between the reciprocal of the droplet radius and pressure difference across the two-phase interface are presented in Figure~\ref{fig:surfacetension} for the two selected settings. It is clear that the pressure difference does proportionally increase with reciprocal of the droplet radius, which confirms the Laplace's law. The slope of the increase in each case, is the surface tension in the lattice unit generated by the proposed MCMP PP LB model for the chosen two-component, two-phase system under different saturated pressures, and they are 0.1289 at $p = 200$ psia and 0.1112 at $p = 240$ psia. Using the results from different domain sizes, we perform a spatial convergence study for the generated surface tension $\sigma$. The relative errors are computed by comparing with the surface tension generated from the highest grid resolution $nx = ny = 800$. As shown in Figure~\ref{fig:surfacetensionconvergencerate}, the surface tension generated from the proposed MCMP PP LB model is roughly second-order accurate in space. 

At last, the dynamic performance of the proposed MCMP PP LB model is examined. Here we adopt the case of elliptical droplet oscillation under its surface tension force. This test case was widely performance in the past to validate the dynamic performance of SCMP PP LB models~\cite{li2013lattice,xu2015three,kharmiani2019alternative}. The elliptical droplet is created by having an initial cell-volume density distribution as
\begin{equation}
     \bar{\rho}_{i}\left(x,y,t = 0\right) =  \dfrac{\bar{\rho}_{i,l}+\bar{\rho}_{i,v}}{2} + \dfrac{\bar{\rho}_{i,l}-\bar{\rho}_{i,v}}{2}\tanh{\left[\dfrac{2\left(\sqrt{\left(x-x_c\right)^2+\left(y - y_c\right)^2/h^2}-R_{0}\right)}{W}\right]},
     \label{eq:ellipsedroplet}
\end{equation}
where $h\ne 1$ and in this test, it is set to 0.9.

\begin{figure}
\centerline
{\includegraphics[width=90mm]{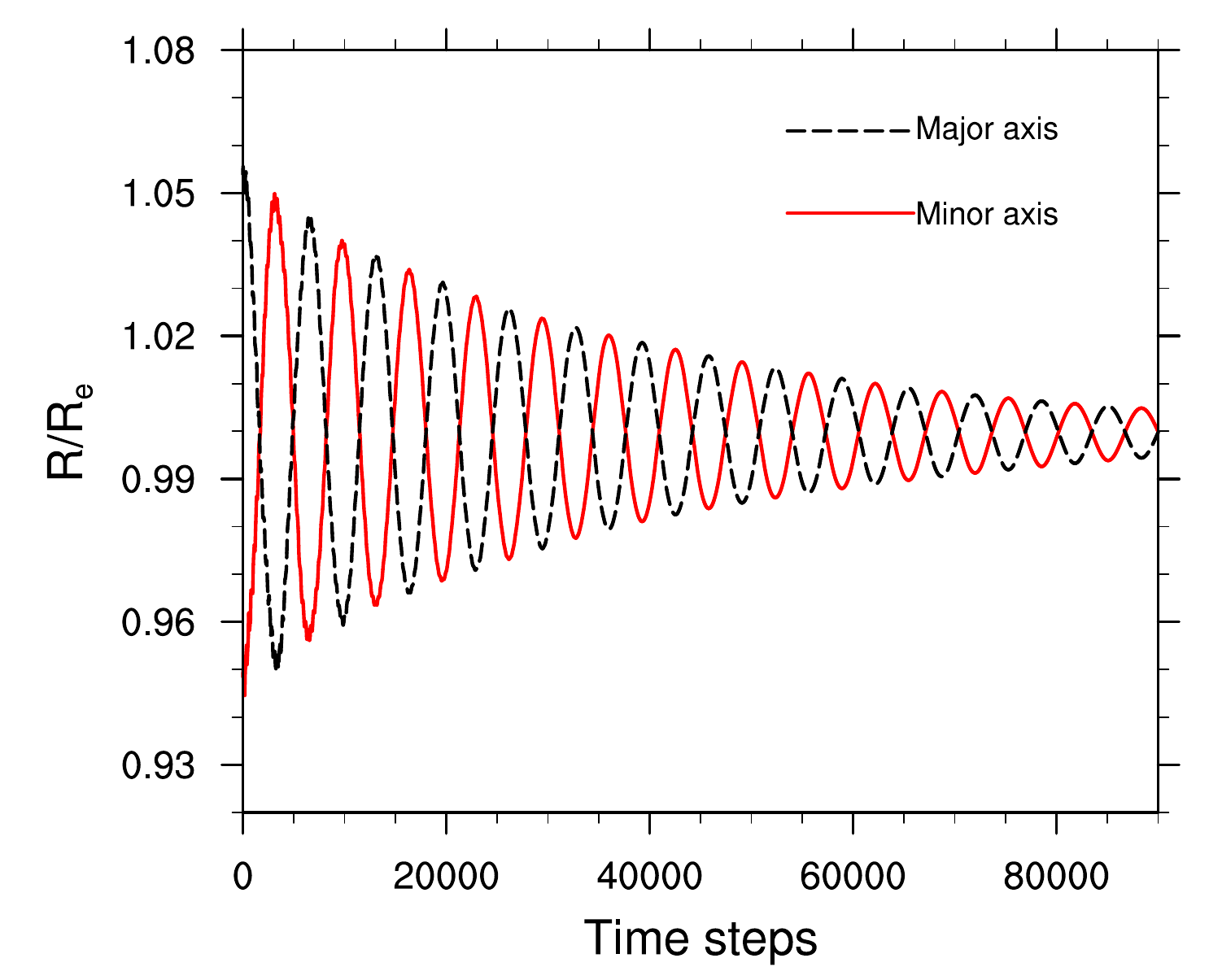}}
\caption{The major and minor axes of the oscillating elliptical droplet as functions of lattice time.}
\label{fig:axis}
\end{figure}

The simulation is again conducted with different grid sizes. The grid size varies from $nx = ny = 200$ to $nx = ny = 800$, and the initial major axis of the initial elliptical droplet changes accordingly as $R_0 = 0.15 nx$. To reduce the effect of the environmental fluids around the droplet, for the $C_3$ and $nC_5$ hydrocarbon mixture, we reduce the prevailing temperature to $T = 580^{\circ}R$, and lower the saturated pressure to $p = 70$ psia, which creates an enhanced phase density ratio about 58. Under this thermodynamic condition, the optimal value of $\beta$ to ensure thermodynamic consistency in the droplet case changes to $1.815$. In Figure~\ref{fig:axis}, the evolution of the major and minor axis with lattice time is presented for the case with $nx = ny = 400$. It is clear that the elliptical droplet keeps oscillating from its equilibrium circular shape. The reducing amplitude of the oscillation is due to the viscous dissipation. For the oscillation motion to persist longer, the kinematic viscosity of the fluid has been reduce to a relatively small value $\nu = 0.03$. This leads to a relaxation time $\tau = 0.59$ in the simulations. 

For this case, we mainly compare the oscillation period, which has a classic theoretical solution~\cite{lamb1932hydrodynamics} 
\begin{equation}
    T_a = 2\pi \left[\frac{n\left(n^2 - 1\right)\sigma}{\rho_l R_e^3}\right]
    \label{eq:theoreticalperoid}
\end{equation}
where $n = 2$ is the mode for initial elliptical droplet shape, $R_e$ is the equilibrium radius of the droplet, which is computed as $R_e = \sqrt{R_{max} R_{min}}$, with $R_{max} = R_0$ and $R_{min} = R_0 h$. With different grid resolutions, the resulting oscillation periods are compared with the theoretical values, and this comparison is tabulated in Table~\ref{tab:periods}. The differences between the numerical and theoretical results mainly come from two sources, the existence of a vapor phase around the droplet, which is neglected in the derivation of Eq.~(\ref{eq:theoreticalperoid}), and the finite interface thickness. With the proposed MCMP PP LB model, the thickness of diffused interface remains constant in lattice length under a given thermodynamic condition. Therefore, when the grid size is expanded, the region of the interface would occupy a smaller section of physical domain, which leads to more accurate prediction of the oscillation period. This is confirmed by the reducing relative errors with larger grid sizes in Table~\ref{tab:periods}.   

\begin{table*}
\caption{Comparison between theoretical and numerical periods of oscillating elliptical droplet. }
\footnotesize
\begin{center}
\begin{tabular}
{c c c c c c c c c c c}
\hline
\toprule
$T_a$& $nx = 200$&$nx = 300$&$nx = 400$&$nx = 600$&$nx = 800$\\
\midrule 
Theory (Eq.~(\ref{eq:theoreticalperoid}))& 2129 & 4011 & 6165 & 11310 & 17403\\
Proposed model& 2407 & 4308 & 6548 & 11861 & 18112\\
Error& 9.86\% & 7.40\% &6.21\% & 4.87\% & 4.08\%\\
\bottomrule
\end{tabular} 
\end{center}
\label{tab:periods}
\end{table*}

\section{Concluding and remarks}\label{sec:conclusion}

In this work, we pointed out the thermodynamic inconsistencies found in MCMP PP LB models used for immiscible fluids and the potential pitfalls involved in applying those models to predict phase behavior of miscible fluids, such as hydrocarbon mixtures. In order to be consistent with MC thermodynamics, thermodynamic information must be incorporated to LB models based on phase information, rather than for individual components.
Following this philosophy, a MCMP PP LB model that can predict the phase behavior of hydrocarbon mixtures should be designed via defining the Shan-Chen force for the whole phase with a thermodynamically meaningful MC EOS, and then non-uniformly distributing this total force to each component via a properly designed force split factor associated with the volatility of the component. The only prior attempt of force splitting in a MCMP PP LB model was proposed by Gong {\it et al}.~\cite{gong2014lattice}. However, in Gong {\it et al.}'s pioneer study, the force split factor was designed empirically through trial and error with no clear physical justification. In this study, we derived a mathematical expression for the force split factor that is consistent with how the total Shan-Chen force was defined in MCMP PP LB model. This force split factor allows the MCMP PP LB model to reproduce phase densities and molar composition consistent with MC thermodynamics, which is controlled by the iso-fugacity rule.

We validated the proposed model using four cases, a limiting case of two components being identical water, three cases of hydrocarbon mixtures of propane and pentane. In the limiting case, we showed that the MCMP PP LB models designed for immiscible components failed to reproduce the phase behavior of pure water, which confirmed their thermodynamic inconsistency. With the case of realistic hydrocarbon mixtures, we demonstrated that Gong {\it et al}.'s force split factor could not predict the correct molar composition in both phases. The present MCMP PP LB model, on the other hand, ensured that the resulting phase densities and molar fractions were qualitatively consistent with equilibrium thermodynamics. Full thermodynamic consistency, however, also relied on the use of Kupershtokh {\it et al.}~\cite{kupershtokh2009equations}'s definition with a properly tuned parameter $\beta$ to compute the total Shan-Chen force. To complete the validation, we also provided benchmark tests in cases with more general curved interface. It had been shown that the proposed MCMP could confirm the Laplace's law and correctly predict the oscillation period of an elliptical droplet under the surface tension force.

{\bf Acknowledgements}: Funding support from Energi Simulation and the William A. Fustos Family Professorship in Energy and Mineral Engineering at Penn State University is gratefully acknowledged. Authors would like to thank Mr. Zhicheng Wang at Penn State University for generating the phase envelope in Figure~\ref{fig:phaseenvelope} and proofreading the manuscript. Authors are also grateful to Dr. Jingwei Huang at Texas A\&M University who notified the authors about the relevant study in Ref.~\cite{gong2014lattice}.

\bibliography{mybibfile}

\end{document}